\documentclass[paper]{JHEP3}
\usepackage{amsmath}
\usepackage{amssymb}
\usepackage{latexsym}
\usepackage{graphicx}
\usepackage{psfrag,fancyhdr,epsfig}

\newcommand{\Tco}{\mathcal{T}}
\newcommand{\as}{a_{d}}
\newcommand{\rs}{r_{d}}
\newcommand{\oms}{\omega_{d}}
\newcommand{\Ain}{A^{\text{in}}_{\Lambda}}
\newcommand{\Aout}{A^{\text{out}}_{\Lambda}}
\newcommand{\Bin}{B^{\text{in}}_{\Lambda}}
\newcommand{\Bout}{B^{\text{out}}_{\Lambda}}

\newcommand{\tort}{r_\ast}

\newcommand{\expct}[3][]{\left\langle#3\left|#2\right|#3\right\rangle_{\mathrm{#1}}}
\newcommand{\ket}[1]{\left.\left|#1\right.\right\rangle}

\title{Bulk emission of scalars by a rotating black hole}

\author{Marc Casals and Sam Dolan\\
School of Mathematical Sciences, University College Dublin,
Belfield, Dublin 4, Ireland. \\
E-mail: \email{Marc.Casals@ucd.ie}, \email{Sam.Dolan@ucd.ie}}

\author{Panagiota Kanti\\
Division of Theoretical Physics, Department of Physics, University
of Ioannina, Ioannina GR-451 10, Greece \\
E-mail: \email{pkanti@cc.uoi.gr}}

\author{Elizabeth Winstanley\\
School of Mathematics and Statistics, The University of Sheffield, Hicks Building,\\
Hounsfield Road, Sheffield S3 7RH, United Kingdom.
\\ E-mail: \email{E.Winstanley@sheffield.ac.uk}}

\abstract{
We study in detail the scalar-field Hawking radiation
emitted into the bulk by a higher-dimensional, rotating black
hole. We numerically compute the angular eigenvalues, and solve
the radial equation of motion in order to find transmission
factors. The latter are found to be enhanced by the angular
momentum of the black hole, and to exhibit the well-known effect
of superradiance. The corresponding power spectra for scalar
fields show an enhancement with the number of dimensions, as in
the non-rotating case. On the other hand, the proportion of the
total (i.e., bulk+brane) power that is emitted into the bulk
decreases monotonically with the angular momentum. We compute the
total mass loss rate of the black hole for a variety of black-hole
angular momenta and bulk dimensions, and find that, in all cases,
the bulk emission remains significantly smaller than the brane
emission. The angular-momentum loss rate is also computed and
found to have a smaller value in the bulk than on the brane.
}


\preprint{}

\begin{document}

\section{Introduction}
Theories with Large Extra Dimensions \cite{ADD, RS} have proven
attractive to theorists in recent years. Of particular interest
are models in which matter is confined to a four-dimensional
hypersurface (the \emph{brane}), but gravity (and possibly scalar
fields) is free to propagate in a higher-dimensional compact space
(the \emph{bulk}). Phenomenologically, such models are intriguing
because they appear to resolve the so-called hierarchy problem. In
other words, they explain why gravity is observed to be so much
weaker (on the macroscale) than the other forces.

In these models, the fundamental energy scale of gravity
($M_\ast$) is related to the Planck energy ($M_P$) by
$M_\ast^{2+n} \sim M_P^2 R^{-n}$, where $R$ and $n$ are the size
and number of extra dimensions.  Hence, $M_\ast$ may be many
orders of magnitude lower than the Planck energy $M_{P}$. This
raises the tantalising prospect that black holes may be created
through `trans-Planckian' particle collisions \cite{creation}. It
has been suggested that $M_\ast$ may be probed by high-energy
cosmic ray collisions \cite{cosmic}, or the next generation of
particle accelerators \cite{colliders}. If created, mini black
holes would evaporate rapidly through emitting Hawking radiation
\cite{hawking}. Experimental detection of Hawking emission would
provide a clear signal of black hole creation. Accurate
measurement of the power spectrum would enable us to deduce the
properties of the underlying spacetime itself. For these reasons,
accurate theoretical modelling of the emission spectrum has become
a high priority, and it has received much attention in recent
years.

Black holes are thought to undergo four stages of evaporation:
the so-called `balding', `spin-down', `Schwarzschild', and
`Planck' phases \cite{Kanti, reviews}. Thanks to a number of
analytical \cite{kmr1, Frolov1, kmr2} and numerical \cite{HK1,
graviton-schw} studies, emission from the Schwarzschild phase is
now well described. Attention has recently shifted to emission in
the rotating phase \cite{Frolov2, DHKW, CKW, CDKW}.

In this paper, we conduct a numerical study of scalar emission in
the bulk by a rotating higher-dimensional black hole. Our work
complements a range of existing studies of the `spin-down' phase
of black hole evolution. In previous papers in this series we have
examined on-the-brane emission of scalars \cite{DHKW}, fermions
\cite{CKW} and photons \cite{CDKW} by a rotating black hole. In
parallel, other groups have pursued complementary lines of inquiry
\cite{IOP}. The bulk emission of scalars from the rotating phase
has received increasing attention. Analytic approximations for the
greybody factors (transmission factors) were recently derived
\cite{CEKT}. The analytic work is complemented by two recent
numerical studies \cite{jung-park-2005, kobayashi} of bulk scalar
emission from 5D  and 6D rotating holes. Here, we present
comprehensive new exact results for bulk dimensionalities $n =
1,2, \dots 6$, and a range of black hole angular momenta.

This study aims to investigate the claim that \emph{black holes
radiate mainly on the brane} \cite{emparan}.  Studies of the
Schwarzschild phase of evolution found that graviton
\cite{graviton-schw} and scalar \cite{HK1} emission in the bulk is
small compared with emission of standard model fields on the
brane, due chiefly to the multiplicity of standard model particle
species. Hence, only a small fraction of the black hole energy is
`lost' in the bulk. It is widely suspected, but unproven, that the
same remains true when the black hole is rotating. As the amount
of energy emitted in the bulk inevitably defines the one on the
brane -- and its observable signatures, this point should be duly
clarified. Two species of particles are usually assumed to
propagate in the bulk, gravitons and scalars. Whilst it is
possible \cite{Kodama} to derive a set of
partial differential equations (PDEs)
describing the
gravitational perturbations of a higher-dimensional rotating black
hole, it is not yet clear whether the PDEs can be decoupled into
ordinary differential equations (ODEs)
for all modes (specifically, the scalar and vector modes)
when the black hole has a single angular momentum parameter\footnote{In
the case of equal angular momenta, the ODEs have been
separated for some modes - see \cite{KLR} and \cite{MS}.}. On the
other hand, in the case of scalars, the decoupling of PDEs is
indeed realised, and a comprehensive study of their emission can
be performed. Apart from the importance in its own right of the
scalar emission, the study of the scalar field in the bulk can
also offer a qualitative understanding of higher-dimensional
graviton emission from a rotating black hole. In the sections that
follow, we compute the strength of scalar emission in brane and
bulk channels in the rotating phase, and calculate the proportion
of the scalar energy which enters the bulk. We make the
assumptions of a minimally-coupled field, negligible brane
tension, and a black hole with just one non-zero angular momentum
component.

The paper is structured as follows. In section {\ref{sec-theory}}
we outline the relevant theory, viz.~a mathematical description of
a rotating black hole in a bulk spacetime; the decomposition of
the scalar field into radial and angular parts; and the relevant
formulae for Hawking emission. In section \ref{sec-nummeth} the
numerical methods employed to calculate angular eigenvalues and
transmission factors are described. In section \ref{sec-results}
we present our numerical results. We start with a brief review of
the Schwarzschild phase results, before moving on to consider the
rotating phase. The effect of superradiance on the transmission
factors is examined. We present brane and bulk power spectra for a
range of $n$ and black hole angular momenta $a$. We compute the
total scalar emission rate, and the fraction which is radiated
into the bulk. We also consider the loss of angular momentum from
the spinning hole. Finally, in section \ref{sec-conclusions} we
conclude with a discussion of the physical implications of our
work.

\section{Theoretical background\label{sec-theory}}
The well-known Myers-Perry solution \cite{MP} describes the
gravitational field of a ($4+n$)-dimensional uncharged rotating
black hole. We will confine our attention to black holes created
by the collision of particles on the brane. Hence we assume that
the black hole metric has only one non-zero angular momentum
component, in a plane parallel to the brane. The line-element
takes the form
\begin{align}
\label{metric}
ds^2 =& -\left( 1 - \frac{\mu}{\Sigma r^{n-1}} \right) dt^2 -
 \frac{2 a \mu \sin^2 \theta}{\Sigma r^{n-1}} dt \, d\psi +
 \frac{\Sigma}{\Delta} dr^2 + \Sigma  \, d\theta^2  \nonumber \\
& + \left( r^2 + a^2 + \frac{a^2 \mu \sin^2 \theta}{\Sigma r^{n-1}} \right)
\sin^2 \theta \, d \psi^2 + r^2 \cos^2 \theta \, d\Omega^2_n,
\end{align}
where
\begin{equation}
\label{sigma}
\Delta = r^2 + a^2 - \frac{\mu}{r^{n-1}}, \quad \quad \Sigma = r^2 + a^2 \cos^2 \theta .
\end{equation}
Here,
%\begin{equation}
\begin{align}
%d\Omega^2_n(\theta_1, \theta_2, \ldots, \theta_{n-1}, \phi)=
%d\theta_1^2+\sin^2\theta_1\left(d\theta_2^2+\sin^2\theta_2(\dots d\phi^2)\dots\right)
& \! \! \! \! \! \! \! \! d\Omega^2_n(\theta_1, \theta_2, \ldots, \theta_{n-1}, \phi)=
\nonumber  \\
%d\theta_1^2+\sin^2\theta_1\left(d\theta_2^2+\sin^2\theta_2(\dots + \sin^2\theta_{n-1}\, d\phi^2) \ldots \right)
& d\theta_{n-1}^2+\sin^2\theta_{n-1}\bigg(d\theta_{n-2}^2+\sin^2\theta_{n-2}\Big(\ldots + \sin^2\theta_2
\left(\ d\theta_1^2+\sin\theta_1^2 d\phi^2\right) \ldots \Big) \bigg)
%\end{equation}
\end{align}
is the line-element on a unit $n$-sphere.
In equation (\ref{metric}), the angle $\psi $ is the azimuthal angle round the axis of rotation of the black hole
in the brane (for four-dimensional black holes, this angle is usually denoted $\phi $).
The mass  $M_{BH}$ of the black hole and its angular momentum $J$ are proportional to $\mu$ and $a\mu$, respectively,
\begin{equation}
M_{BH} = \frac{(n+2) A_{n+2}}{16 \pi G} \mu , \quad \quad J = \frac{2}{n+2} M_{BH} a ,
\label{amudef}
\end{equation}
where $A_{n+2} = 2\pi^{(n+3)/2} / \Gamma[(n+3)/2]$ is the area of
an $(n+2)$-dimensional unit sphere,  and $G$ is the
$(4+n)$-dimensional version of Newton's constant.
Equation (\ref{amudef}) effectively defines the parameters $\mu $ and $a$ which appear in the metric (\ref{metric}).

The radius of the black hole's event horizon, $r_h$, is the
largest, positive root of $\Delta(r) = 0$. The horizon radius
$r_h$ can be employed to define a useful set of dimensionless
variables, which we denote with the subscript $d$,
\begin{equation}
\rs = r / r_h, \quad \quad \as = a / r_h , \quad \quad \oms = \omega r_h. \quad \quad
\end{equation}
%Note that in this paper $\rs$ does not denote the `tortoise' co-ordinate, as is customary.
For $n \ge  1$, there is one unique root in the region $r > 0$, and it can be obtained from the implicit equation
\begin{equation}
 r_h^{n+1}=\frac{\mu}{1+\as^2}.
\end{equation}
For $n>1$ this root exists for any value of $\as$. However, it has
been shown \cite{Harris} that the maximum value of $\as$ for black
holes created by particle collision is $\as^{\text{max}} = n/2 +
1$.

\subsection{Scalar field equations}
In this paper, we study the Hawking emission of scalar fields into
the higher-dimensional `bulk' spacetime.  Therefore we must first
consider the equation of motion for a massless scalar field $\Phi$
propagating in the bulk, with minimal coupling to the geometry,
which satisfies the field equation:
\begin{equation} \label{ScalarODE}
\frac{1}{\sqrt{-g}} \partial_\mu \left(\sqrt{-g} \, g^{\mu \nu} \partial_\nu \Phi\right) = 0.
\end{equation}
Here, $\sqrt{-g}$ is the determinant of the metric,
\begin{equation}
\sqrt{-g} = \Sigma r^n \sin \theta  \cos^n \theta \prod_{i=1}^{n-1} \sin^i \theta_i .
\end{equation}
%{\bf (I know this is the formula in eq.5~\cite{CEKT}, but based on the sphere line element in p.337~\cite{MP} should it not be with something like $\prod_{i=1}^{n-1} \sin^{n-i} \theta_i$ instead?)}
%
Substituting in (\ref{ScalarODE}) the factorization ansatz
\begin{equation}
\label{variable separation}
\Phi_{\Lambda} = e^{-i \omega t} e^{i m \psi} R_{\Lambda}(r) S_{\Lambda}(\theta)
Y_{jn}( \theta_1, \ldots, \theta_{n-1}, \phi) ,
\end{equation}
where $\Lambda\equiv \left\{l,m,j,\omega\right\}$,
yields three coupled second-order equations. The first equation,
\begin{equation}
\sum_{k=1}^{n-1} \frac{1}{\Pi^{n-1}_{i=1} \sin^i \theta_i} \partial_{\theta_k}
\left[ \left( \prod_{i=1}^{n-1} \sin^i \theta_i \right)
\frac{ \partial_{\theta_k} Y_{jn} }{ \prod_{i>k}^{n-1} \sin^2 \theta_i}  \right]
+ \frac{ \partial_{\phi\phi} Y_{j n}}{\prod_{i=1}^{n-1} \sin^2 \theta_i} + j(j+n-1) Y_{jn} = 0  ,
\end{equation}
defines the \emph{hyperspherical harmonics} on the $n$-sphere, $Y_{jn}$ \cite{Muller}.
The remaining two equations determine the angular $S_{\Lambda}(\theta)$ and radial $R_{\Lambda}(r)$
mode functions. They are \cite{CEKT}:
\begin{equation}
\frac{1}{\sin\theta \cos^n \theta} \partial_\theta
(\sin \theta \cos^n \theta \, \partial_\theta S_{\Lambda})
+ \left( \omega^2 a^2 \cos^2 \theta - \frac{m^2}{\sin^2 \theta}
- \frac{j(j+n-1)}{\cos^2 \theta} + E_{\Lambda}  \right) S_{\Lambda} = 0,
\label{eq-angular}
\end{equation}
and
\begin{equation}
\frac{1}{r^n} \partial_r \left( r^n \Delta \partial_r R_{\Lambda} \right)
+ \left( \frac{K^2}{\Delta} - \frac{j(j+n-1)a^2}{r^2} - \left[E_{\Lambda}
+ 2am\omega - a^2 \omega^2\right] \right) R_{\Lambda} = 0,
\label{eq-radial}
\end{equation}
where $K = (r^2 + a^2) \omega - am$.
The angular (\ref{eq-angular}) and radial (\ref{eq-radial})
equations are coupled through the angular eigenvalue
$E_{\Lambda}$, which depends on the spheroidicity parameter
$a\omega$. There is no known closed analytic form for the
eigenvalues, but at least three methods are available for their
calculation. First, and simplest, $E_{\Lambda}$ may be expressed
as a power series in $a \omega$. The coefficients of the power
series up to 4th order in $a \omega$ are given in~\cite{BCC}.
Alternative numerical methods are Leaver's method of continued
fractions \cite{Leaver}, and a numerical matching method,
described fully in section \ref{subsec-angular}.

Note that the angular equation (\ref{eq-angular}) together with
the boundary conditions of regularity of the solution at the
points $\theta=0$ and $\theta=\pi/2$ constitutes a
Sturm-Liouville eigenvalue problem. Orthogonality of the solutions
with the appropriate measure follows, and we normalize them so
that
\begin{equation}
\int_0^{\pi/2}d\theta\sin\theta\cos^n\theta S_{l'mj\omega}S_{lmj\omega}=\delta_{l'l}.
\label{eq-orthogonal-angular}
\end{equation}
Similarly, the hyperspherical harmonics are orthogonal and we normalize them as
\begin{equation}
\int d\phi\prod_{i=1}^{n-1}d\theta_i  \prod_{k=1}^{n-1} \sin^k \theta_k  Y_{j'n}Y_{jn}^{*}=\delta_{j'j}.
\label{eq-orthogonal-hyperspherical}
\end{equation}

\subsection{Hawking emission}

The quantization of the scalar field in the higher-dimensional
curved space-time follows through in essentially the same manner
as in the 4-D case. We will therefore merely indicate the main
steps in the quantization and derivation of the flux formulae and
we refer the interested reader to, for example,
papers~\cite{Ott&Winst,Cas&OttB,F&T} for the 4-D case and to
\cite{Frolov2} for the $n=1$  case.

We define the inner product of two solutions $\Psi$ and $\Phi$ of the scalar equation (\ref{ScalarODE})
in the bulk as
\begin{equation}
\langle\Psi,\Phi\rangle=\frac{i}{2}\int_{\Sigma}d\Sigma^{\mu}\sqrt{-g}
\left(\Phi_{;\mu}^*\Psi-\Psi^*_{;\mu}\Phi\right),
\end{equation}
where a semicolon denotes the covariant derivative, $\Sigma$ is a
complete Cauchy surface and $d\Sigma^{\mu}$ is a future-directed
normal to $\Sigma$ (in this section only, the symbol $\Sigma$ will
be used to denote a complete Cauchy surface rather than the
function in equation (\ref{sigma})). Using the scalar field equation it
can be checked that this inner product is independent of the
choice of the complete Cauchy surface $\Sigma$.

We will follow Myers and Perry~\cite{MP}, and define null coordinates $u$ and $v$ as
\begin{equation}
u\equiv t-\tort, \qquad v\equiv t+\tort,
\end{equation}
where $\tort$ is a ``tortoise'' radial co-ordinate defined as
\begin{equation}
\tort\equiv r + \frac{1}{2\kappa} \log{\left(\frac{r-r_h}{r_h}\right)} ,
\end{equation}
and $\kappa$ is the surface gravity, which is given by~\cite{MP},
\begin{equation}
\kappa \equiv \frac{(n+1)+(n-1)\as^2}{2 (1+\as^2)r_h}= \lim_{r \rightarrow r_h}
\left[\frac{\Delta}{2 (r_h^2+a^2)(r-r_h)}\right].
\end{equation}
The null congruences are tangent to surfaces of constant $u$ or $v$ at the horizon ($r=r_h$) and at
infinity ($r\to \infty$), but this is generally not the case for other values of $r$.
We can also define the Kruskal-like co-ordinates,
\begin{equation}
U \equiv -e^{-u \kappa}, \qquad
V \equiv e^{ v \kappa}.
\end{equation}
We will choose $\Sigma=\mathcal{J}^-\cup\mathcal{H}^-$ as the Cauchy surface, where
$\mathcal{J}^-$ is the past null infinity given by $U=-\infty$
and $\mathcal{H}^-$ is the past horizon given by $V=0$.
The surface element in these regions can be calculated,
\begin{subequations}
\label{surface}
\begin{align}
d\Sigma^{\mu}&=  \frac{(r_h^2+a^2)(r-r_h)}{(r^2 + a^2 \cos^2 \theta)  U}l^{\mu}dU \, d\theta \,
d\psi  \, d\theta_1\dots d\theta_{n-1}
\, d\phi  \quad \text{at}\quad \mathcal{H}^- ,
\\
d\Sigma^{\mu}&=  -\frac{1}{2\kappa V} n^{\mu}dV \, d\theta \, d\psi \, d\theta_1\dots d\theta_{n-1}
\, d\phi  \quad \text{at}\quad \mathcal{J}^-,
\end{align}
\end{subequations}
where $l$ and $n$ are two null vector fields,
\begin{subequations}
\begin{align}
l&=l^{\mu}\frac{\partial}{\partial x^{\mu}}=
\frac{(r^2+a^2)}{\Delta}\left[\frac{\partial}{\partial t}+ \frac{a}{(r^2+a^2)}\frac{\partial}{\partial \psi}\right]
-\frac{\partial}{\partial r},
\\
n&=n^{\mu}\frac{\partial}{\partial x^{\mu}}=
\frac{(r^2+a^2)}{\Delta}\left[\frac{\partial}{\partial t}+ \frac{a}{(r^2+a^2)}\frac{\partial}{\partial \psi}\right]
+\frac{\partial}{\partial r}.
\end{align}
\end{subequations}

The radial equation (\ref{eq-radial}) has singular points at $r =
r_h$ and infinity.  We can define two sets of linearly-independent
radial solutions (valid for $n > 0$) with the asymptotic
behaviours,
\begin{subequations}
\label{eq:Rin/up}
\begin{align}
 R^{\text{in} }_{\Lambda}   & \sim
\begin{cases}
 (\rs - 1)^{- i \tilde{\omega}/2\kappa},
& \quad r\rightarrow r_+ , \\
 \frac{1}{\rs^{1 + n/2}} \left( \Ain e^{- i \oms \rs} +  \Aout e^{+ i \oms \rs} \right) ,
 & \quad r\rightarrow +\infty ,
\label{ingoing-bc}
\end{cases}
\\
R^{\text{up}}_{\Lambda} & \sim
\begin{cases}
\Bin (\rs - 1)^{- i \tilde{\omega}/2\kappa}+ \Bout (\rs - 1)^{+ i \tilde{\omega}/2\kappa} ,
&
 r\rightarrow r_+ , \\
 \frac{1}{\rs^{1 + n/2}}  e^{+ i \oms \rs}  ,
 &  r\rightarrow +\infty ,
\end{cases}
\end{align}
\end{subequations}
where $\Ain$, $\Aout$, $\Bin$ and $\Bout$ are complex constants, $\tilde{\omega}\equiv \omega - m\Omega_h$
and $\Omega_h = a / (r_h^2 + a^2)$
 is the angular speed of the horizon.
We define the `in' solutions for $\omega>0$ and the `up' solutions for
$\tilde{\omega}>0$.

The field mode solutions $\Phi^{\text{in/up}}_{\Lambda}$ constructed via equation (\ref{variable separation})
from equations (\ref{eq:Rin/up})
satisfy the normalization conditions
\begin{equation}
\label{normalization}
\langle \Phi^
{{\bullet}^*}_{\Lambda},\Phi
^{\bullet'}
_{\Lambda'}\rangle =0, \qquad
\langle \Phi_{\Lambda}
^{\bullet}
,\Phi_{\Lambda'}
^{\bullet'}
\rangle =\delta_{\Lambda\Lambda'}
\delta_{\bullet\bullet'},
\end{equation}
after an appropriate normalization factor is included, where
symbols $\bullet$ and $\bullet'$ denote either the `in' or the
`up' solutions. The normalization factors can be calculated using
the surface elements (\ref{surface}) and, when included, the field
modes are explicitly given by
\begin{align}
\Phi^{\text{in}}_{\Lambda}&=
\frac{1}{2\pi \sqrt{\omega} r_h^{n/2+1}|\Ain|}
e^{-i \omega t} e^{i m \psi} R^{\text{in}}_{\Lambda}(r) S_{\Lambda}(\theta)
Y_{jn}( \theta_1, \ldots, \theta_{n-1}, \phi) ,
\\
\Phi^{\text{up}}_{\Lambda} &=
\frac{1}{2\pi \sqrt{K(r_h)} r_h^{n/2}|\Bout|}
e^{-i \omega t} e^{i m \psi} R^{\text{up}}_{\Lambda}(r) S_{\Lambda}(\theta)
Y_{jn}( \theta_1, \ldots, \theta_{n-1}, \phi) ,
\end{align}
where the orthogonality relations (\ref{eq-orthogonal-angular})
and (\ref{eq-orthogonal-hyperspherical}) have been used.

The field mode solutions `in' and `up' form a complete set of orthonormal solutions to the scalar equation
(\ref{ScalarODE}) in the outer region of the space-time.
A general solution of the scalar equation can therefore be written as a linear combination
of this basis
\begin{equation}
\Phi(x)=\sum_{\Lambda} \sum_{\bullet=\text{in,up}}
\left[
a_{\Lambda}\Phi^{\bullet}_{\Lambda}(x)+
a_{\Lambda}^{*}\Phi^{\bullet^*}_{\Lambda}(x)
\right],
\end{equation}
where the integration over the frequency is for $\omega>0$ for the `in' modes
and for $\tilde{\omega}>0$ for the `up' modes.
Promoting the fields and the Fourier coefficients $a_{\Lambda}$ to operators $\hat{a}_{\Lambda}$,
using the normalization conditions (\ref{normalization}) and
imposing the standard canonical commutation relations for the field and its canonical momenta leads to the
commutation relations
\begin{equation}
\left[\hat{a}_{\Lambda}
^{\bullet}
,\hat{a}_{\Lambda'}
^{\bullet'}
\right]=
\left[\hat{a}
^{\bullet^{\dagger}}_{\Lambda},\hat{a}
^{\bullet'^{\dagger}}_{\Lambda'}\right]=0,
\qquad
\left[\hat{a}_{\Lambda}
^{\bullet}
,\hat{a}
^{\bullet'^{\dagger}}_{\Lambda'}\right]=\delta_{\Lambda\Lambda'}
\delta_{\bullet\bullet'} \, .
\end{equation}
In the eternal version of the space-time (i.e. in the full
analytic extension of the Myers-Perry metric) we can construct
from the field modes $\Phi^{\text{in/up}}_{\Lambda}$ a complete
orthonormal basis of modes which are positive-frequency with
respect to both the affine parameter $U$ on the past horizon
$\mathcal{H}^-$ and the affine parameter $t$ on past null
infinity $\mathcal{J}^-$. The past Unruh state $\ket{U^-}$ is
defined \cite{Ott&Winst,Unruh'76} as the quantum state which is
annihilated by the Fourier coefficients -- as operators -- of this
basis of modes.
%Such a state models the state of a black hole at late times.
This state models an evaporating black hole and is therefore the relevant one for computing the Hawking emission.

The stress-energy tensor for a minimally coupled scalar field is given by
\begin{equation} \label{SET}
T_{\mu\nu}\left[\Phi,\Phi^*\right]=\frac{1}{2}\left[\Phi_{;\mu}\Phi_{;\nu}^*+\Phi_{;\nu}\Phi_{;\mu}^*\right]
-\frac{1}{2}g_{\mu\nu}\Phi^{;\alpha}\Phi_{;\alpha}^*.
\end{equation}
It can be proven \cite{Ott&Winst,Cas&OttB}
that the expectation value of the stress-energy tensor operator when the scalar
field is in the past Unruh state is given by
\begin{equation} \label{SET,Unruh}
\expct{\hat{T}_{\mu\nu}}{U^-}
=
\sum_{l,j,m}
\Bigg(
\int_0^{\infty}d{\tilde{\omega}}\,
\coth\left(\frac
{\tilde{\omega}}{2T_{{H}}}
\right)
T_{\mu\nu}\left[{}\Phi_{\Lambda}^{\text{up}},{}\Phi_{\Lambda}^{\text{up}^*}\right]
+
\int_0^{\infty}d{\omega}\,
T_{\mu\nu}\left[{}\Phi_{\Lambda}^{\text{in}},{}\Phi_{\Lambda}^{\text{in}^*}\right]
\Bigg)
\end{equation}
with the Hawking temperature
\begin{equation}
T_H = \frac{\kappa}{  2 \pi} .  \label{eq-T-hawking}
\end{equation}
It is well-known that the expectation value of the stress-energy
tensor is generally divergent and it must be renormalized
appropriately. One method of renormalization is the technique of
covariant geodesic point separation. In this approach, the pair of
fields appearing in each quadratic term in the stress-energy
tensor (\ref{SET}) are evaluated at separate points, $x$ and
$x^\prime$.  This defines a bi-tensor $T_{\mu\nu}(x,x')$. The
expectation value of this bi-tensor is found for the desired
quantum state, and certain purely geometric terms
$T_{\mu\nu}^{\text{div}}(x,x')$ are subtracted from it. Finally,
the two points are moved together again, i.e. the coincidence
limit $x'\to x$ is taken. In 4-D,
Christensen~\cite{Christ'76,Christ'78} calculated the geometric
subtraction terms using the Schwinger-DeWitt
expansion~\cite{DeWitt,Schwinger} for the Hadamard elementary
function. The Schwinger-DeWitt expansion for the non-rotating,
higher-dimensional case is given in \cite{FMP}. In principle this
could be used to compute the geometric subtraction terms in higher
dimensions, following \cite{Christ'76,Christ'78}, but fortunately
we do not require their precise form in our analysis.
%Fortunately, it is easy to check that the Schwinger-DeWitt expansion and the geometric subtraction terms continue to be valid in the case of the higher-dimensional space-time metric (\ref{metric}).

In~\cite{F&T} it was shown that those components of the
unrenormalized expectation value of the stress-energy tensor with
one index either $t$ or $\psi$ and the other index $r$ (as in
(\ref{fluxes}) below) are not divergent and so do not require
renormalization. The proof is based on two properties. Firstly, we
use the symmetry of the metric under $(t,\psi)\to (-t,-\psi)$.
Secondly, each of the geometric subtraction terms contains an even
number of covariant derivatives $\sigma^{\mu}$ of the biscalar of
geodetic interval~\cite{Synge} after an average is taken over a
separation in the $\sigma^{\mu}$ and $-\sigma^{\mu}$ directions.
By choosing a point-splitting in the radial direction, these two
properties then ensure that the geometric subtraction terms
$T_{tr}^{\text{div}}(x,x')$ and $T_{r\psi }^{\text{div}}(x,x')$
vanish (see \cite{F&T} for the full argument). Even without
explicitly computing the geometric subtraction terms, it can be
shown that these two properties remain true in higher dimensions,
so that the argument of \cite{F&T} follows through, and we have
$T_{tr}^{\text{div}}(x,x')=T_{r\psi }^{\text{div}}(x,x')=0$
independent of $n$, the number of extra dimensions.
%It is relatively straightforward to check that these two properties remain true in the higher-dimensional case of the Myers-Perry metric, and the proof follows through. %, and so these components require no renormalization.

Having established that the $tr$ and $\psi r$ components do not
require renormalization, the fluxes of energy and angular momentum
are given by
\begin{align} \label{fluxes}
\frac{d E}{dt} & =\Delta\int_Sd\Omega_{n+2} \expct{\hat{T}_{tr}}{U^-}, \nonumber \\
\frac{d J}{dt} & =\Delta\int_Sd\Omega_{n+2} \expct{\hat{T}_{r}^{\psi }}{U^-},
\end{align}
where $S$ is any surface of constant $t$ and $r$. The total rate
of energy emission of massless scalars into the bulk is found by
taking a sum over all angular modes. Taking for the surface $S$ in
(\ref{fluxes}) a $(2+n)$-sphere boundary at infinity and using
equations (\ref{SET,Unruh}) and (\ref{eq:Rin/up}), it can be shown
\cite{Frolov2, CEKT} that
\begin{equation}
\frac{d^2 E}{dt d\omega} = \frac{1}{2\pi} \sum_{l, j, m} \frac{\omega}{\exp(\tilde{\omega} / T_H) - 1}
\, N_j \Tco_{\Lambda}  \label{eq-power-emission}
\end{equation}
and the emission rate of angular momentum into the bulk is given by
\begin{equation}
\frac{d^2 J}{dt d\omega} = \frac{1}{2\pi} \sum_{l, j, m} \frac{m}{\exp(\tilde{\omega} / T_H) - 1}
\, N_j \Tco_{\Lambda} . \label{eq-am}
\end{equation}
Here, $\Tco_{\Lambda}$ is the energy-dependent \emph{transmission
factor} (also known as `greybody factor') and is given by the
ratio of the ingoing and outgoing fluxes,
\begin{equation}
\Tco_{\Lambda} = 1 - \left| \frac{\Aout}{\Ain}  \right|^2 .   \label{eq-trans}
\end{equation}
In addition, $N_j$ is a degeneracy factor
(given in equation (11) of \cite{Muller})
which accounts for the multiplicity of modes in the bulk,
\begin{equation}
N_j = \frac{(2j + n - 1) (j + n - 2)!}{j! (n - 1)!} .  \label{eq-degeneracy}
\end{equation}
All parts of equation (\ref{eq-power-emission}) can be determined
analytically, except for the transmission factors
$\Tco_{\Lambda}$. Qualitatively, the transmission factor is a
dimensionless measure of the proportion of a given mode that can
escape from the black hole horizon to reach infinity. Hence,
$\Tco_{\Lambda} = 1$ corresponds to `total' emission, and
$\Tco_{\Lambda} = 0$ to `total' back-reflection of Hawking
radiation. The transmission factors in turn determine the shape
and magnitude of the power spectrum. Approximate expressions for
$\Tco_{\Lambda}$ were recently presented in the literature
\cite{CEKT}. In this study, we determine them numerically by
solving the radial equation (\ref{eq-radial}). Our numerical
method is outlined in section \ref{subsec-numerical-trans}.
%It is straightforward to compute the fluxes (\ref{eq-power-emission}) and (\ref{eq-am}) once we have determined the transmission coefficients.

\section{Numerical methods\label{sec-nummeth}}
In this section we briefly describe the numerical methods we
employ for computing the angular eigenvalues and the transmission
factors.

\subsection{Angular eigenvalues\label{subsec-angular}}

The allowed values of the angular momentum quantum numbers $l$, $j$ and $m$ are restricted \cite{BCC}.
The azimuthal number $m$ can take any integer value, and $l$ and $j$ are positive or zero integers, such that
\begin{equation}
l \ge j + |m|   \quad \quad \text{and} \quad \quad l - j - |m| = 2 k , \quad \text{where}\ k \in \{0, \mathbb{Z}^+\}.
\label{condition nums.}
\end{equation}
The second equation in (\ref{condition nums.}) implies that only
even [odd] values of $l$ are allowed when $(j+|m|)$ is even [odd].
Note that the angular equation (\ref{eq-angular}) has two regular
singular points at $\theta=0$ and $\theta = \pi$,  just like the
4-D spin-weighted spheroidal equation (equation (\ref{eq-angular}) with
$j=n=0$), plus an extra regular singular point at $\theta=\pi/2$,
which is a non-singular point in the 4-D case. The presence of
this extra regular singular point means that half the regular
solutions in the 4-D case will now, for $n>0$, become irregular.
This is the reason why the angular momentum quantum number $l$ is
forced to have the same parity as $j+|m|$ (as opposed to being
allowed to be any non-negative integer in the 4-D case). Note also
that the whole of the physical region in the bulk is covered by
the regime $\theta\in [0,\pi/2]$. However, when restricting
ourselves to the brane (by taking $\theta_1=0$), the whole
physical region is covered by extending the regime to $\theta\in
[0,\pi]$ (in the same way as when vertically slicing a 2-sphere
into two halves we must extend the regime of the polar angle from
$\theta\in[0,\pi]$ to $\theta\in[0,2\pi]$ in order to cover the
full circle).

The numerical method we used for solving the angular equation in the bulk, equation (\ref{eq-angular}),
is essentially the same -- with the obvious modifications --
as the shooting method for the 4-D case as described in detail in \cite{Cas&OttA},
to which we refer the interested reader.
We require power series expansions around the two integration end-points $\theta=0$ and $\theta=\pi/2$.
Near the end-point $\theta=\pi/2$ we used the power series expansion given in equation (3.4) of \cite{BCC}.
Near the other end-point $\theta=0$ (or $\theta=\pi$, by symmetry) we insert the expansion
\begin{equation}
S_\Lambda=\left(\sin\theta\right)^{|m|}\left(\cos\theta\right)^j\sum_{p=0}^{\infty}a_p\left(1\pm \cos\theta\right)^p
\end{equation}
into equation (\ref{eq-angular}) and obtain the recursion relation:
\begin{equation}
%\begin{aligned}
%a_{i+1}&=\frac{1}{(2(i+1)(i+|m|)+1)}\Big\{
%((\left[3(i-1)+2(n+2j)+4(|m|+1)\right]i-\delta)-c2-y(3))a_i+
%\\
%&
%(-((i-1)(i+n+2j+2|m|)-\delta)+3c2+y(3))a_{i-1}+
%-3c2a_{i-2}+c2a_{i-3})
%\Big\}
a_{p+1}=\frac{1}{\alpha_p}\left\{\beta_pa_p+\gamma_pa_{p-1}+\delta_pa_{p-2}+\epsilon_pa_{p-3}\right\} ,
%\end{aligned}
\end{equation}
where
\begin{equation}
\begin{aligned}
\alpha_p&=2(p+1)(p+|m|+1),\\
\beta_p&=\left[3(p-1)+2(n+2j)+4(|m|+1)\right]p-(a\omega)^2-E_{\Lambda}+j(j+n+1)\\
& \, \, \, \, \, \, \, \, +|m|(n+|m|+2j+1),\\
\gamma_p&=3(a\omega)^2-p(p+n+2j+2|m|-1)+n+E_{\Lambda}-j(j+n-1)\\
& \, \, \, \, \, \, \, \,-|m|(n+|m|+2j-1),\\
\delta_p&=-3(a\omega)^2,\\
\epsilon_p&=(a\omega)^2.
\end{aligned}
\end{equation}
It is easy to check that for $n=j=0$ this recursion relation
coincides with the corresponding one in the 4-D case (see, e.g., equations (2.5)--(2.8) with spin $s=0$
in \cite{BCC}).

The eigenvalues $E_{\Lambda}$ in the 4-D case become degenerate for large $a\omega$ \cite{Cas&OttA}.
Fortunately, this degeneracy does not occur in the bulk case \cite{BCC}.
Therefore, the integration of the angular equation with our technique
is more amenable in the bulk than in 4-D for large $a\omega$.

\subsection{Transmission factors\label{subsec-numerical-trans}}

To compute the transmission factors $\Tco_{\Lambda}$ we solved the
radial equation (\ref{eq-radial}) numerically.  We started with
the ingoing solution close to the horizon and integrated out to
large $\rs$. Here, we matched the numerical solution onto the
asymptotic form (\ref{ingoing-bc}) to determine $\Ain$ and $\Aout$
and hence, via (\ref{eq-trans}), $\Tco_{\Lambda}$. Typically, we
integrated from $\rs = 1.001$ to a matching point at $\rs = 100$.

The solution close to the horizon may be found using the method of
Frobenius.  The radial equation (\ref{eq-radial}) may be rewritten
\begin{equation}
\frac{d^2 R_{\Lambda}}{d \eta^2} = \frac{A(\eta)}{\eta} \, \frac{d R_{\Lambda}}{d \eta}
+ \frac{B(\eta)}{\eta^2} \, R_{\Lambda} ,
\end{equation}
where $\eta = r / r_h - 1$. Here, $A(\eta)$ and $B(\eta)$ are functions with Maclaurin series expansions,
\begin{equation}
A(\eta) = \sum_{j = 0} A_j \, \eta^j , \quad \quad \quad B(\eta) = \sum_{j=0} B_j \, \eta^j .
\end{equation}
The coefficients $A_j$ and $B_j$ are easily determined using a mathematical software package such as Maple.
Near the horizon, the radial solution may be expressed as a power series in $\eta$,
\begin{equation}
R_{\Lambda}(\eta) = \eta^s \, \sum_{j = 0} R_j \, \eta^j .
\end{equation}
The index $s$ satisfies the indicial equation $s(s-1) - s A_0 -
B_0 = 0$, which has solutions $s = \pm i
\tilde{\omega}/(2\kappa)$. The negative root is the correct choice
for an ingoing solution.  The coefficients $R_j$ are determined
from the recurrence relation
\begin{equation}
\left[ (s+j)(s+j-1) - (s+j)A_0 - B_0 \right] R_j = \sum_{k=1}^{j} \left[ (s+j-k)A_k + B_k \right] R_{j - k} .
\end{equation}
We used this method to compute expansion coefficients up to $R_5$.

The ingoing and outgoing solutions in the large-$\rs$ limit may be
found by a similar method.  First we make the substitution
\begin{equation}
R_{\Lambda}(r) = \frac{e^{\pm i \oms \rs}}{\rs^{1 + n/2}} \, X_{\Lambda}(r),   \quad \quad \quad n \in \mathbb{Z}^+ .
\end{equation}
The positive (negative) sign in the exponent gives the outgoing
(ingoing) solution.  This substitution leads to a differential
equation which can be written
\begin{equation}
\frac{d^2 X_{\Lambda}}{d z^2} = \left(\frac{\pm 2i \oms}{z^2} +  \frac{C(z)}{z} \right) \, \frac{dX_{\Lambda}}{dz}
+ \frac{D(z)}{z^2} \, X_{\Lambda}
\end{equation}
where $z = 1/\rs$. Again, $C(z)$ and $D(z)$ have Maclaurin series
expansions which can be easily found with a symbolic algebra
package. The radial function $X_{\Lambda}(z)$ also has a Maclaurin
series expansion, $X_{\Lambda}(z) = \sum_{j=0} X_j z^j$, with
coefficients $X_j$ determined by the recurrence relation
\begin{equation}
\pm 2 i \oms (j + 1) X_{j+1} = j (j - 1) X_j - \sum_{k=0}^j ((j - k) C_k + D_k) X_{j - k} .
\end{equation}
We computed coefficients $X_j$ up to tenth order.  Note that the
ingoing solution is simply the complex conjugate of the outgoing
solution.

\section{Results\label{sec-results}}
In this section we present numerical results for the emission of scalars by a rotating
higher-dimensional black hole. Our objective is to compare the relative magnitudes of
emission into the bulk and on to the brane. First, we begin with a review of the results
for the Schwarzschild (non-rotating) phase \cite{HK1}. Next, we examine the effect of
black hole rotation upon the transmission factors and emission spectra. In its
rotating phase, the black hole is shedding both mass and angular momentum; we determine
the loss rates for both quantities. We conclude by presenting bulk-to-brane emission
ratios for a range of values of $\as$ and $n$.

\subsection{\label{subsec:non-rotating}Non-rotating black holes $(\as = 0$)}

A detailed study of scalar brane and bulk emission in the
non-rotating phase was presented in \cite{HK1}.  Here we review
some key results of that work.

Figure \ref{schw-1} shows the power spectrum on the brane (left)
and in the bulk (right) for $n = 1,2 \ldots 6$.   The total
emission rate, or power, is equivalent to the area under the
curves. The total emission depends strongly on the number of bulk
dimensions, $n$. For all $n$, the total power emitted on the brane
exceeds that emitted into the bulk (note the $y$-scales in figure
\ref{schw-1}). This is despite the fact that particles emitted
into the bulk are, on average, more energetic than those emitted
on the brane.

\FIGURE{
\mbox{\includegraphics[width=15cm]{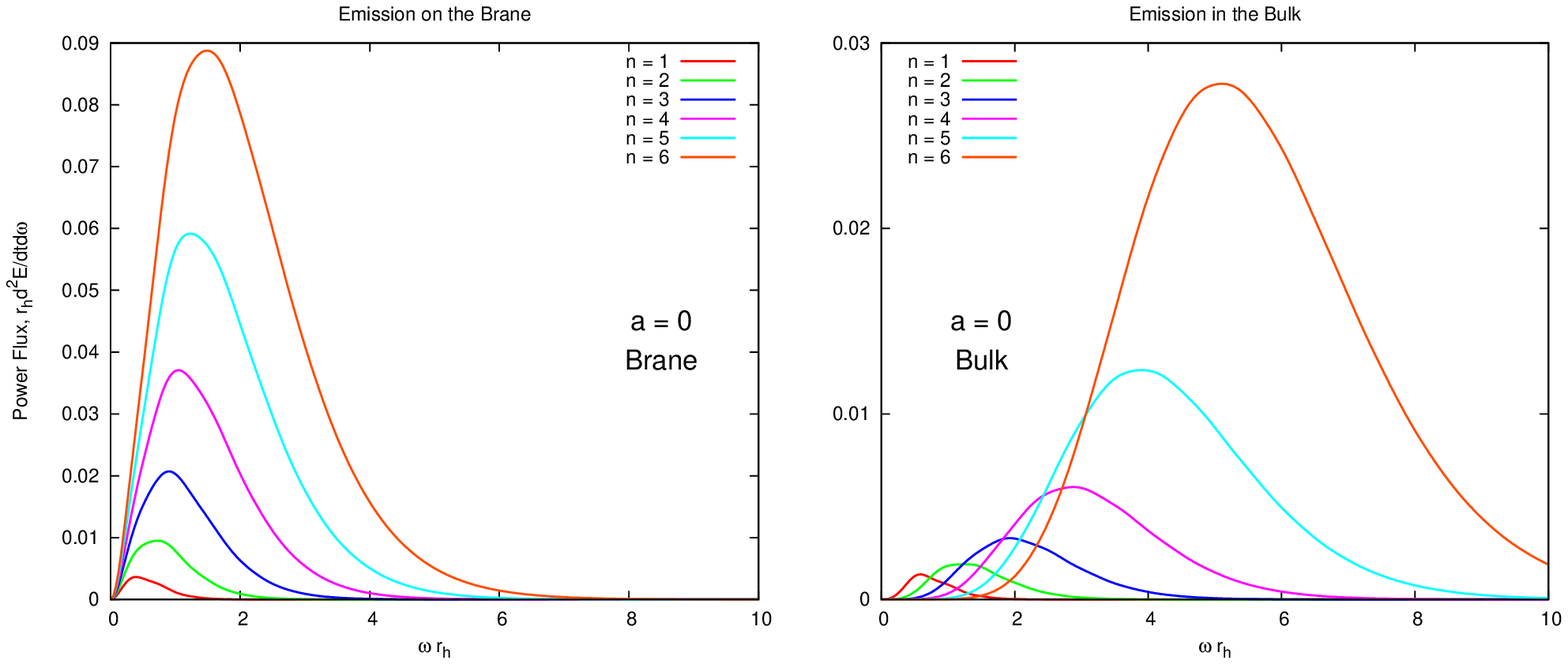}}
\caption{{\emph{Brane and bulk emission from a non-rotating higher-dimensional black hole.}}
The left plot shows the power emitted on the brane, for various numbers of space-time dimensions ($n = 1,2 \ldots 6$).
The right plot shows the power emitted into the bulk. Note the factor of 3 difference in the scales on the $y$-axis. }
\label{schw-1}
}

\TABLE{
%\begin{center}
\begin{tabular}{| l | c c c c c c |}
\hline
 & $\; n = 1  $ & $n = 2 $ & $ n = 3 $ & $ n = 4 $ & $ n = 5 $ & $ n = 6 $ \\
\hline
Total Power, $r_h^2 \, dE/dt$ & $0.00371$ & $0.0134$ & $0.0362$ & $0.0823$ & $0.170$ & $0.339$ \\
\% in bulk & $28.3\%$ & $19.9\%$ & $17.9\%$ & $19.6\%$ & $24.8\%$ & $34.0\%$ \\
\hline
\end{tabular}
%\end{center}
\caption{{\emph{Proportion of scalar power emitted into the bulk by
a non-rotating black hole.}}   These quantities were calculated by
numerically integrating the power spectra up to a frequency cutoff
of $\omega r_h = 10$. The values are in good agreement with
earlier studies \cite{HK1}.} \label{table-ratios-schw}
}

Table \ref{table-ratios-schw} shows the total power (in units of
$1/r_h^{2}$) for $n = 1,2 \ldots 6$.  The total power increases
monotonically with $n$. The proportion of the total power that is
emitted into the bulk (i.e. the `missing' energy fraction)
decreases for $n$ from $1$ to $3$, but increases again for $n >
3$.

\FIGURE{
%\begin{center}
\includegraphics[width=14cm]{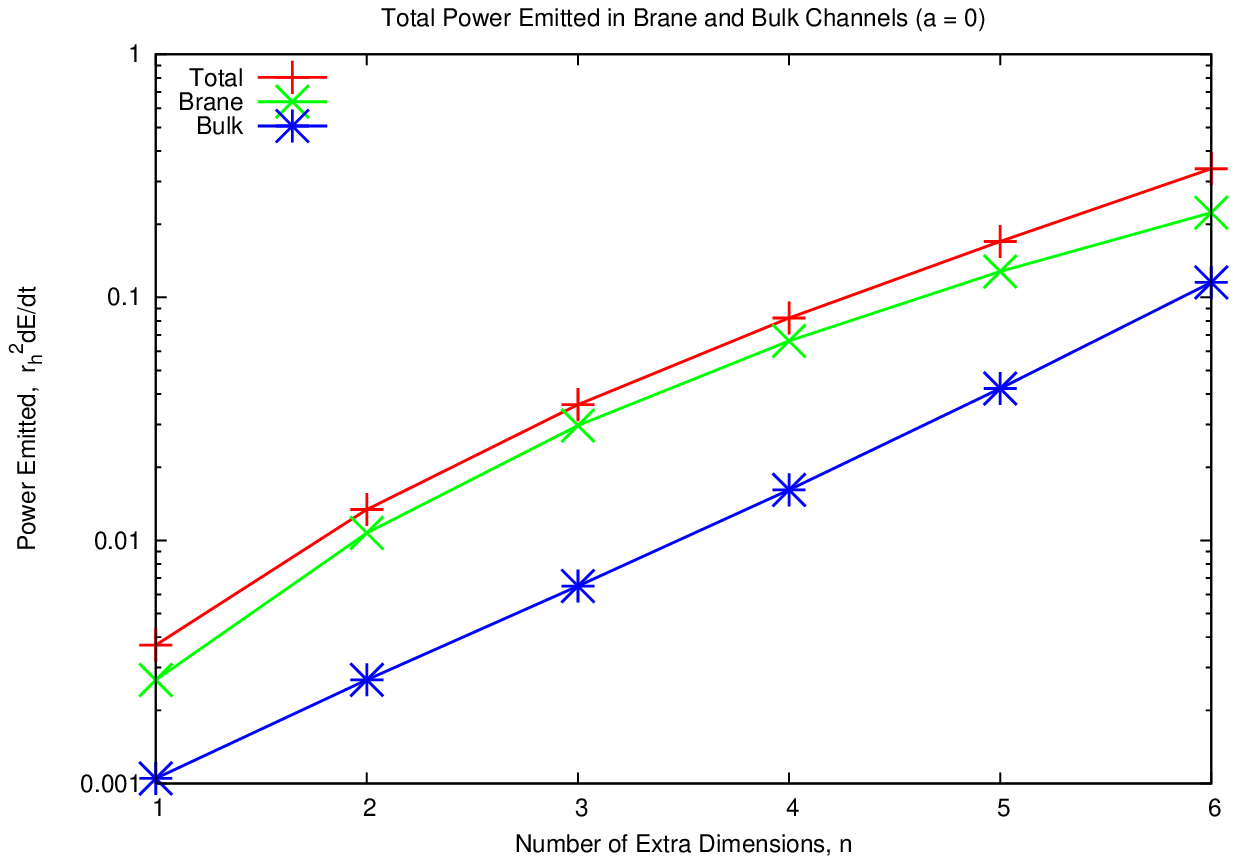}
%\end{center}
\caption{{\emph{Total power (in scalar particles) emitted by a non-rotating black hole.}}
The blue line shows the power emitted in the bulk, and the green line shows the power emitted on to the brane,
for various $n$ (the number of extra dimensions). Note the logarithmic scale on the $y$-axis. }
\label{schw-2}
}

Figure \ref{schw-2} shows how the proportion of energy lost in the
bulk depends on $n$.  The total power emitted in the bulk appears
to obey a power law relationship, $r_h^2 \tfrac{dE}{dt} \sim A
\sigma^n$, where $A \sim 0.0004 $ and $\sigma \sim 2.54$. The
power emitted on the brane increases less rapidly with $n$. Hence
the proportion of total energy lost in the bulk decreases up to $n
= 3$, but increases for $n > 3$.

The key question that we seek to address in the next section is:
how does the proportion of energy lost into the bulk change when
the black hole is rotating? In other words, is the standard
claim~\cite{emparan} that \emph{black holes radiate mainly on the
brane} still correct in the spin-down phase of black hole
evolution?

\subsection{Rotating black holes, $\as > 0$}
Here we consider a black hole rotating in the $\theta = \pi/2$
plane on the brane.  Introducing a single plane of rotation has
many consequences: it modifies the angular eigenvalues, the
transmission factors, and the form of the mode sum. Rotation
breaks the azimuthal degeneracy, so that modes of different $m$
contribute different amounts of power.

Previous studies \cite{DHKW, IOP} have shown that, overall,
rotation enhances the emission on the brane.  Higher angular
momentum modes become increasingly important, and the black hole
emits particles at higher energies. The superradiant effect
increases the emission from the co-rotating modes, particularly
for low $n$. For fast-rotating black holes, the $m = l$ modes
dominate the emission. This leads to an oscillatory power
spectrum, and a rapid loss of black hole angular momentum.

\subsubsection{Transmission factors}
Let us begin by considering the transmission factors $\Tco_{\Lambda}$, defined by
equation (\ref{eq-trans}). An analytic expression for the transmission factor was derived in
\cite{CEKT} under the assumption that both the energy $\omega_d$ and angular
momentum $a_d$ parameters are much lower than unity. For the indicative case of the
$l = j = m = 0$ mode, the dominant behaviour, as $\omega \rightarrow 0$, was also
computed and found to be given by the expression
\begin{equation}
\Tco_{000\omega} \approx \frac{4 \pi (1+ \as^2)^2 (\omega r_h)^{n + 2} }{A_\ast \, 2^n (n+1)
\Gamma^2(\tfrac{n+1}{2}) (2 - D_\ast) } ,
\label{cekt-result}
\end{equation}
where $A_\ast = (n+1) + (n-1) \as^2 $ and $D_\ast = 1 - 4\as^2 /
A_\ast^2$. In figure \ref{fig-creek1}, we compare the approximation
of equation (\ref{cekt-result}) with our numerically-determined
transmission coefficients. We find good agreement in the regime
$\mathcal{T}_{\Lambda} \lesssim 0.1$. It is clear that the effect
of rotation is to shift the $l = j = m = 0$ transmission curve to
lower energies.

\FIGURE{
%\begin{center}
\includegraphics[width=15cm]{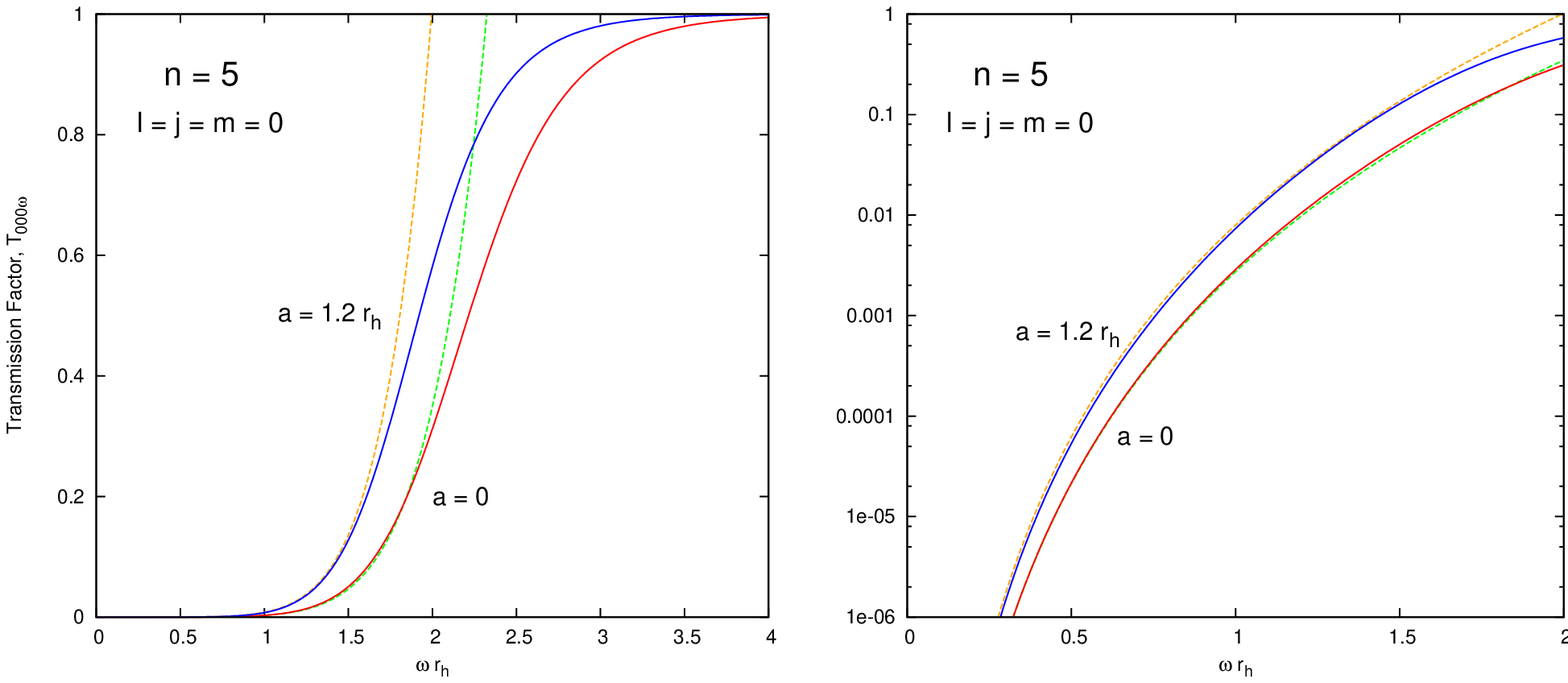}
%\end{center}
\caption{{\emph{Transmission factor of the $l = j = 0$ mode for $n = 5$, for $\as = 0.0$ and $\as = 1.2$.}}
This plot compares the approximation \cite{CEKT} (dotted line)
given in equation (\ref{cekt-result}) with our exact numerical results (solid line).
The right plot shows the same data, but with a logarithmic scale on the $y$-axis.}
\label{fig-creek1}
}

To compute the total cross section, it is not sufficient to
consider only the $l = j = m = 0$ modes.  We must compute a sum
over all angular modes. The modes are labelled by three integers:
$l$, the total angular momentum; $m$, the angular momentum in the
plane of rotation; and $j$, the angular momentum in the bulk.
These integers satisfy the condition (\ref{condition nums.}). If
the black hole is rotating then we must compute transmission
factors for each valid combination of $l$, $j$ and $m$. In
practice, we will truncate the mode sum at some suitable $l =
l_{\text{max}}$. Unsurprisingly, the multiplicity of modes in the
bulk is greater than the multiplicity of modes on the brane. For
example, the number of non-degenerate modes $N(L)$ in the range $0
\le l \le L$ is  $N_{\text{brane}} (L) = (L+1)^2$ on the brane,
but is $N_{\text{bulk}}(L) = L(L^2 - 1)/6$ in the bulk. The
computation time increases commensurately with the number of modes
considered.

Figure \ref{fig-transmissionfactors} shows the transmission
coefficients for the first few modes,  for the particular case of
a slowly-rotating 6D hole ($\as = 0.4$). A hierarchy of effects is
clear. The modes are split firstly on $l$, secondly on $m$ and
finally on $j$. The azimuthal splitting on $m$ increases in
magnitude with rotation parameter $\as$.

\FIGURE{
%\begin{center}
\includegraphics[width=13cm]{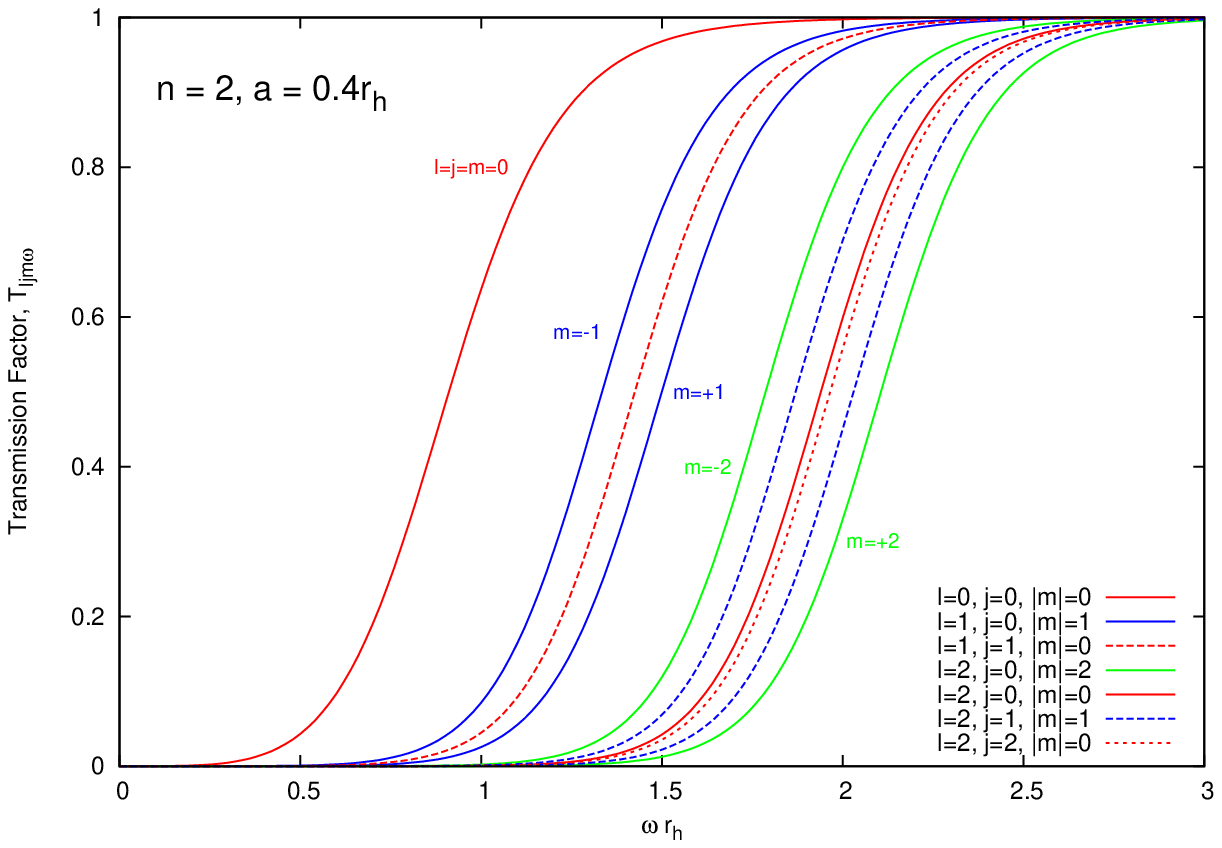}
%\end{center}
\caption{{\emph{Transmission factors for the first few modes, $l = 0$, $1$ \& $2$.}}
The plot shows transmission factors for a 6D black hole rotating at $\as = 0.4$.
The three sets of lines correspond to $l = 0$ (left), $l = 1$ (middle) and $l = 2$ (right).
For each $l$, there is a multiplet of $(l+1)(l+2)/2$ modes.
The modes in each multiplet are split by $m$ and, to a lesser extent, by $j$.}
\label{fig-transmissionfactors}
}

In the frequency regime $\omega < m \Omega_h$,  the superradiance
effect causes the transmission coefficient to become negative.
Nonetheless, the emitted power is positive, since the denominator
of equation (\ref{eq-power-emission}) is also negative in this regime.
Superradiance is most significant for the $m = l$, $j = 0$ modes
(i.e. the maximally-corotating modes). Figure
\ref{fig-superradiance} shows how superradiance affects the $m =
l$  transmission factors of fast-rotating 5D (left) and 6D (right)
black holes. It is clear that superradiance is most significant at
low $n$. In particular, for $n = 1$, superradiance has a
significant effect on the shape of the power spectra, as we see in
the next section.

\FIGURE{
%\begin{center}
\includegraphics[width=15cm]{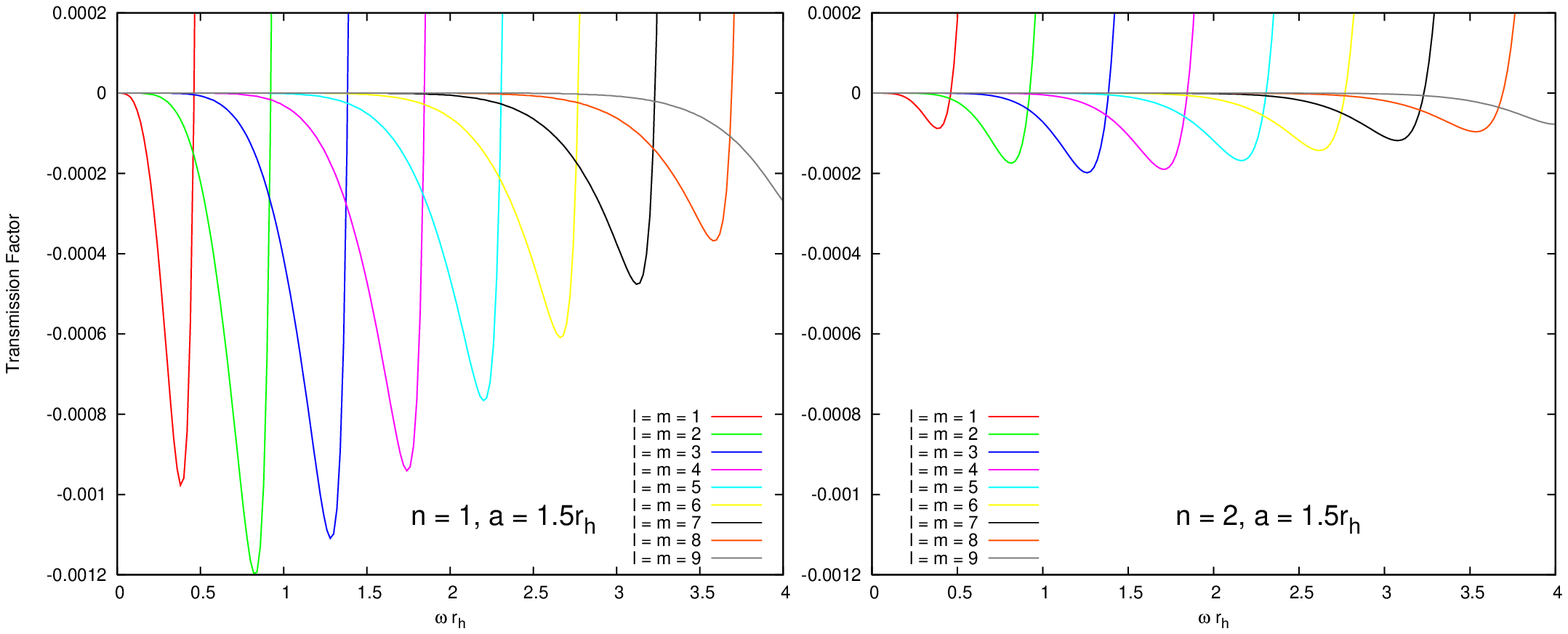}
%\end{center}
\caption{{\emph{Transmission factors for the maximally-rotating $m = l$ modes of a fast-rotating ($\as = 1.5$)
black hole.}}
The left plot shows transmission factors for a 5D ($n=1$) black hole,
and the right plot shows the same modes for a 6D ($n=2$) black hole.
The transmission factors are negative for $\omega < m \Omega_h$ due to the superradiance effect.}
\label{fig-superradiance}
}

\subsubsection{Power spectra}
Let us now consider the effect of rotation on the shape and
magnitude of the spectra of scalar radiation emitted in both the
bulk and the brane. As we saw in the previous section, the shape
and magnitude of the emission spectrum strongly depends on the
number of extra dimensions, $n$. Let us examine two cases in
detail, $n = 2$ and $n = 6$.

\FIGURE{
%\begin{center}
\includegraphics[width=15cm]{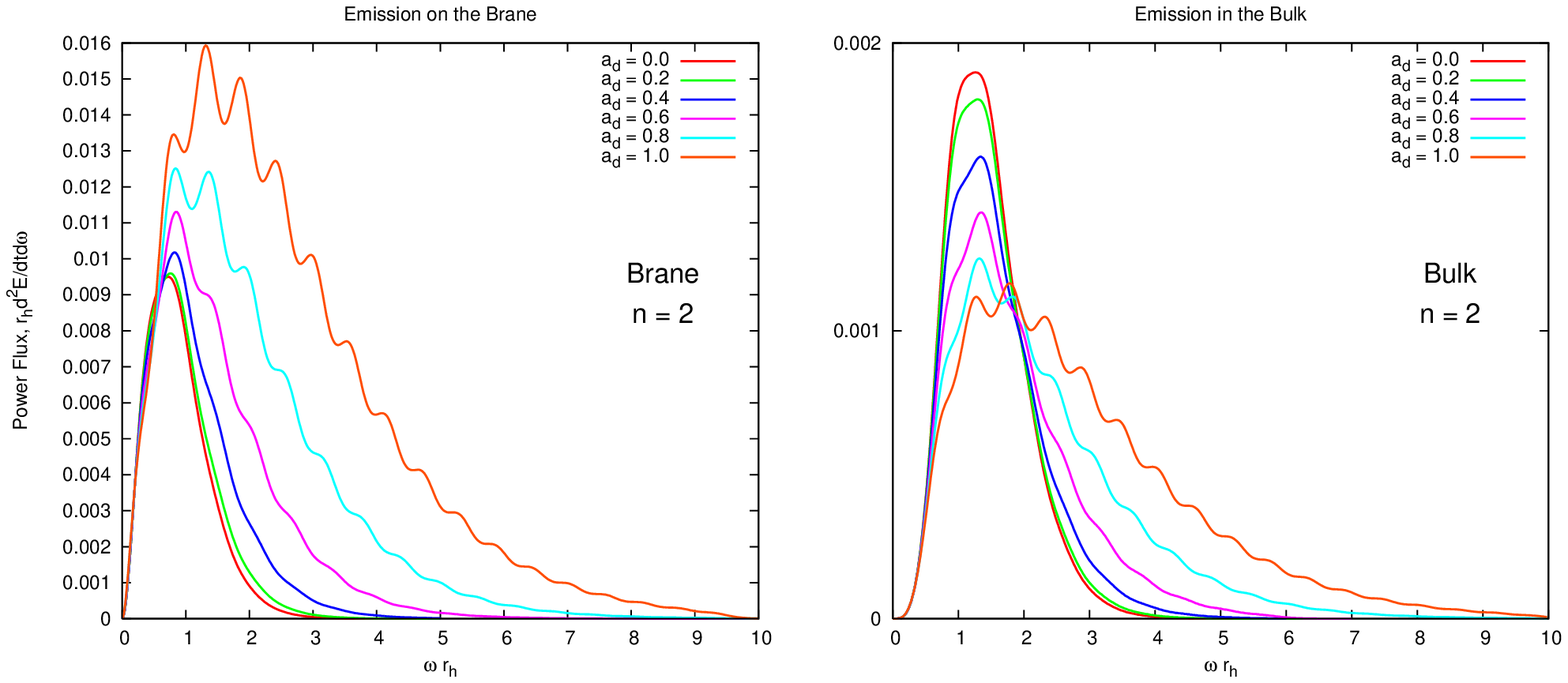}
%\end{center}
\caption{{\emph{Brane and bulk emission from a 6D rotating black hole.}}
The left and right plots show the power emitted on the brane and in the bulk, respectively.
Black hole rotation increases the proportion of the total flux that is emitted on the brane.
Note the order-of-magnitude difference in the scales on the $y$-axis.}
\label{fig-n2-a0to1}
}

Figure \ref{fig-n2-a0to1} shows how rotation changes the emission
spectrum of a 6D ($n = 2$) black hole.  The left panel shows
emission on the brane, and the right panel shows emission in the
bulk. From the left panel, it can be seen that the total emission
on the brane (the area under the curve) increases monotonically
with the rotation parameter $\as$. In the bulk, the situation is
more complicated. A slowly-rotating black hole actually emits less
power into the bulk than a non-rotating hole. However, above $\as
\sim 0.4$, the total emission increases with $\as$. The total
powers for the curves shown in figure \ref{fig-n2-a0to1} are given
in table \ref{table-ratios-n2}. The reduction in bulk emission is
due to the fact that the Hawking temperature (\ref{eq-T-hawking})
decreases monotonically with $\as$. Thus, a rotating black hole is
cooler than a non-rotating hole. On the brane, the reduction in
$T_H$ is more than outweighed by the fact that rotation changes
the transmission factors, and enhances emission in the higher
angular modes. In the bulk, this is not necessarily the case,
particularly at low $\as$.

\TABLE{
%\begin{center}
\begin{tabular}{| l | c c c c c c |}
\hline
 & $\; \as = 0.0 $ & $\as = 0.2 $ & $\as = 0.4  $ & $\as = 0.6 $ &
  $\as = 0.8 $ & $\as = 1.0 $ \\
\hline
Bulk Power & $0.00267 $ & $0.00262$ & $0.00256$ & $0.00266$ & $0.00308$ & $0.00391$ \\
Brane Power & $0.01074 $ & $0.01148$ & $0.01414$ & $0.02003$ & $0.03132$ & $0.05142$ \\
\% in Bulk & $19.9 \%$ & $18.6 \%$ & $15.3 \%$ & $ 11.7\%$ & $9.0 \%$ & $7.1 \%$ \\
\hline
\end{tabular}
%\end{center}
\caption{{\emph{Proportion of scalar power emitted into the bulk by
a 6D rotating black hole.}}  These figures were calculated by numerically
integrating the power spectra, up to a cutoff of $\omega r_h =
10$. The power (mass loss rate) is given in units of $1/r_h^2$.}
\label{table-ratios-n2}
}

In the non-rotating case, emission is primarily in the $l = 0$ and $l =1$ modes. However,
rotation enhances the emission from higher angular modes. The oscillatory pattern in the
emission spectra at higher $\as$ is due to the increasing importance of the superradiant
$m = l$, $j = 0$ modes. This is seen clearly in figure \ref{fig-n2a1_5-modes}, which shows
the contribution of the $m = l$ and $m = l-1$ modes to the overall power spectrum.

\FIGURE{
%\begin{center}
\includegraphics[width=14cm]{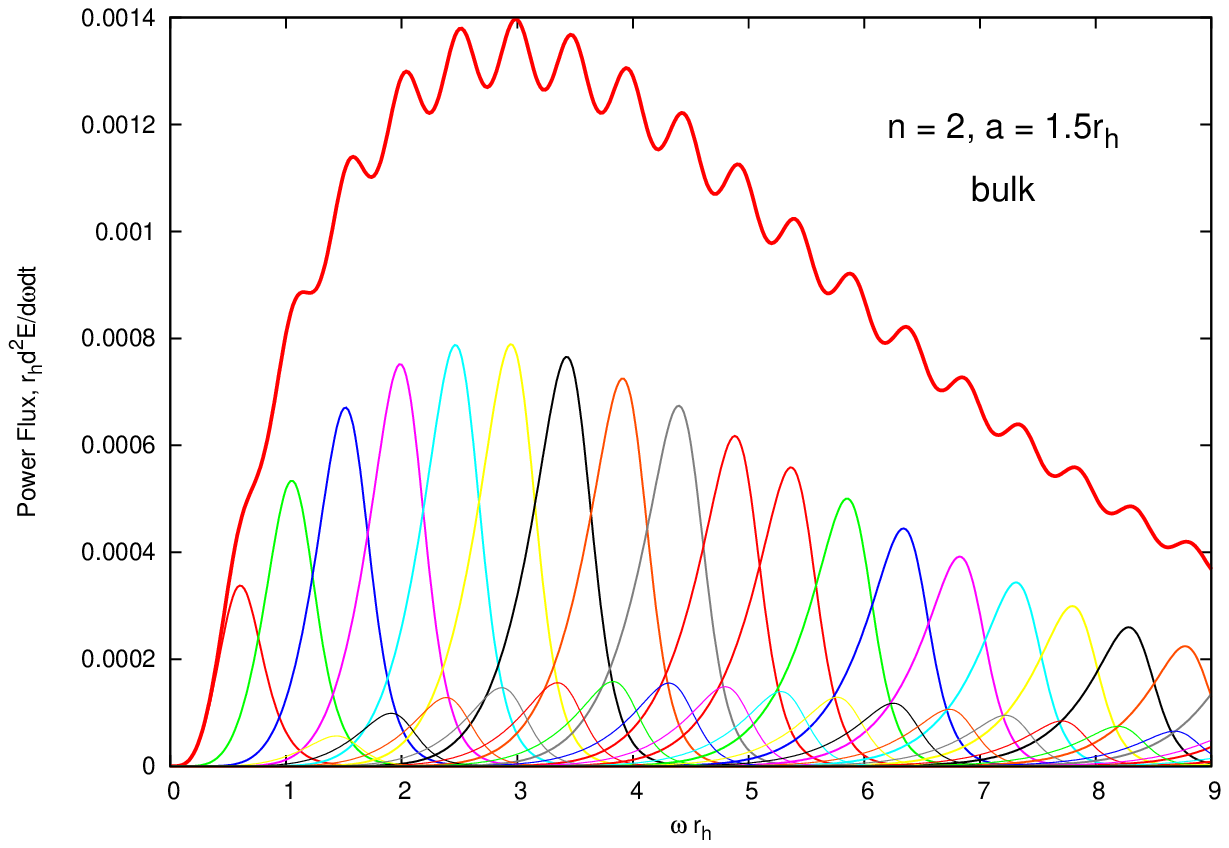}
%\end{center}
\caption{{\emph{Emission from a 6D black hole at $\as = 1.5$.}}
The plot shows that the majority of the emission comes from the maximally-corotating modes ($m = l$).
The $m = l-1$ modes are also plotted.}
\label{fig-n2a1_5-modes}
}

\FIGURE{
%\begin{center}
\includegraphics[width=15cm]{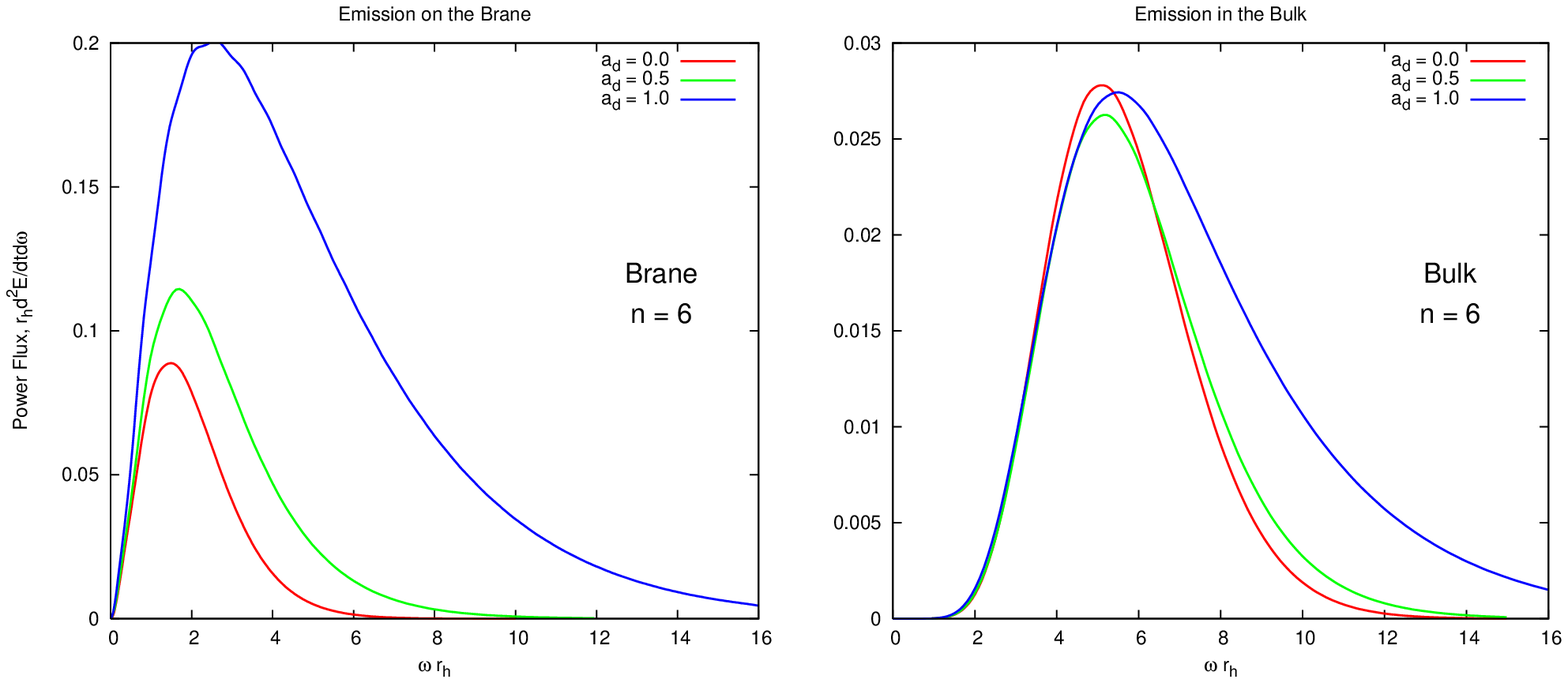}
%\end{center}
\caption{{\emph{Brane and bulk emission from a 10D rotating black hole.}}
The left and right plots show the power emitted on the brane and in the bulk, respectively.
Black hole rotation increases the proportion of the total power that is emitted on the brane.
Note the order-of-magnitude difference in the scales on the $y$-axis.}
\label{fig-n6-a0to1}
}

\FIGURE{
%\begin{center}
\includegraphics[width=14cm]{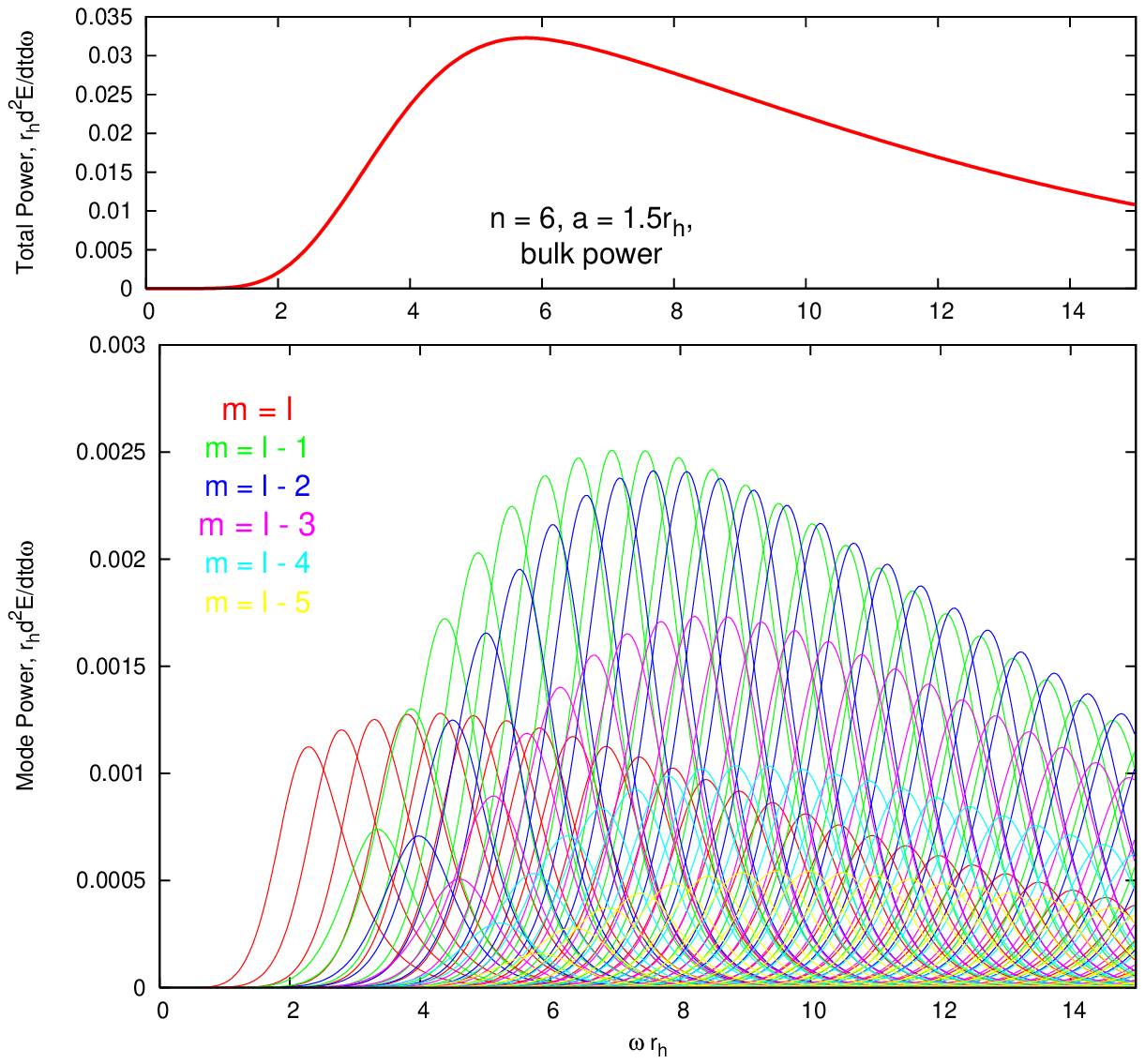}
%\end{center}
\caption{{\emph{Emission from a 10D black hole at $\as = 1.5$.}}
The upper plot shows the total power.
The lower plot shows the contribution from a `forest' of modes, $m = l$ to $m = l-5$.
It can be seen that the $m = l$ modes are no longer dominant at all energies.
Note that a sum over $j$ has been taken where appropriate.}
\label{fig-n6a1_5-modes}
}

% n = 6.
Next, let us consider emission from a 10D ($n = 6$) rotating black
hole.  Figure \ref{fig-n6-a0to1} compares the power spectra on the
brane (left panel) and in the bulk (right panel), at $\as = 0.0$
(red), $\as = 0.5$ (green) and $\as = 1.0$ (blue). It is
immediately clear that rotation enhances emission on the brane
significantly, but has a lesser effect on emission in the bulk. A
large number of angular modes contribute to the total power for $n
= 6$. Figure \ref{fig-n6a1_5-modes} compares the relative
magnitudes of the $m = l$, $m = l-1$ $\ldots$ $, m = l -5$ modes,
at $\as = 1.5$, for $l = 0$, $1,  \ldots 26$. Note that a sum over
$j$ was taken where multiple modes exist with the same $l$ and
$m$. With six bulk dimensions, the degeneracy factor $N_j$
(\ref{eq-degeneracy}) appearing in the mode sum is potentially
very large. This enhances the contribution from the higher $j$
modes. The degeneracy factors increase in the sequence $1$, $n+1$,
$n(n+3)/2$, $(n+5)(n+1)n/6$, $\ldots$ for $j = 0$, $1$, $2$, $3$,
$\ldots$; i.e.  $1$, $7$, $18$, $77$, $\ldots$ for $n = 6$. Modes
close to $m = 0$ have a much higher degeneracy than modes near $m
= l$. For each $m$, the most significant contributions arise from
the $j = l - m$ modes.

\FIGURE{
%\begin{center}
\includegraphics[width=15cm]{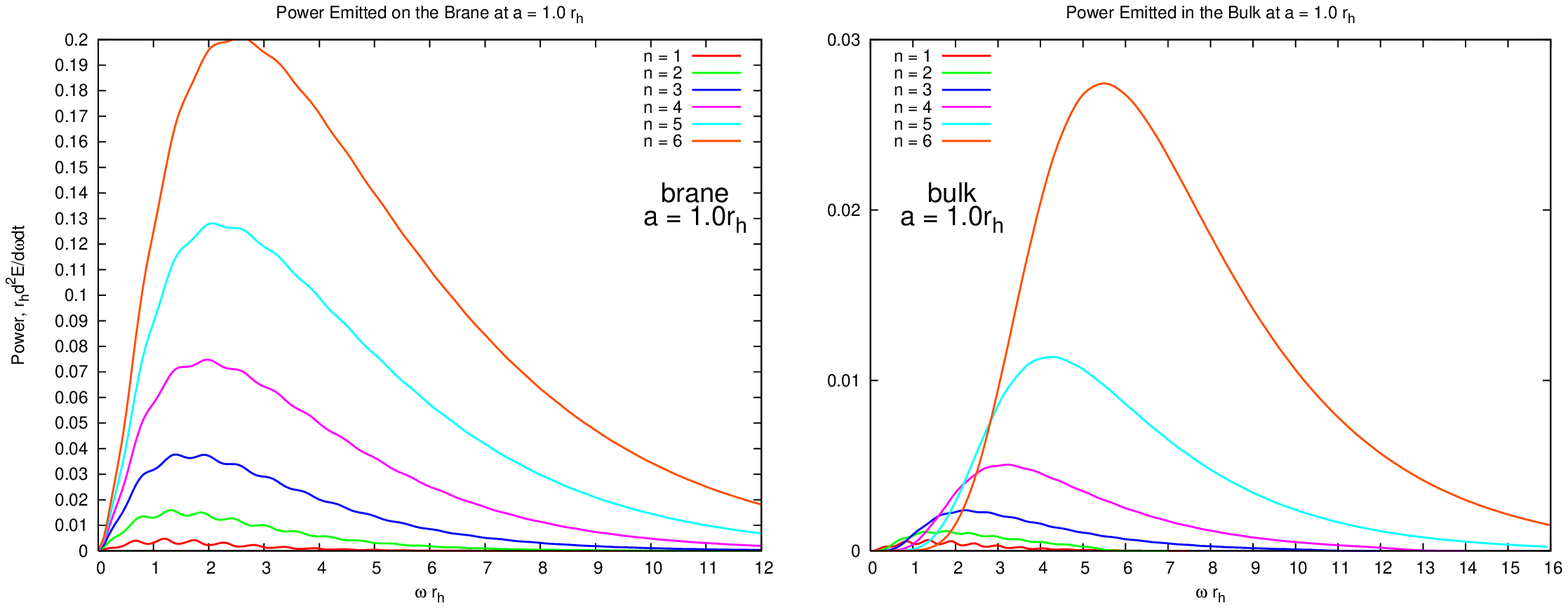}
%\end{center}
\caption{{\emph{Emission by a rotating black hole.}}
These plots compare the power emitted on the brane (left) and in the bulk (right) at $\as = 1.0$.
Note the difference in the scales on the $y$-axis.}
\label{fig-power-all-n}
}

Figure \ref{fig-power-all-n} compares the emission on the brane
(left) and in the bulk (right) at $\as = 1.0$,  for a range of $n
= 1,2 \ldots 6$. Note that the scale on the $y$-axes differs by a
factor of $\sim 7$. Comparing these spectra with the non-rotating
spectra of figure \ref{schw-1}, it is clear that rotation has
reduced the proportion of the overall emission which enters the
bulk. Table \ref{tbl-proportions} lists the fraction of power lost
in the bulk, for a range of scenarios. This data is plotted in
figure \ref{fig-totals}, which shows the \% of scalar power in the
bulk for $\as = 0 \ldots 1$ and $n = 1$, $2, \ldots 6$.

\FIGURE{
%\begin{center}
\includegraphics[width=14cm]{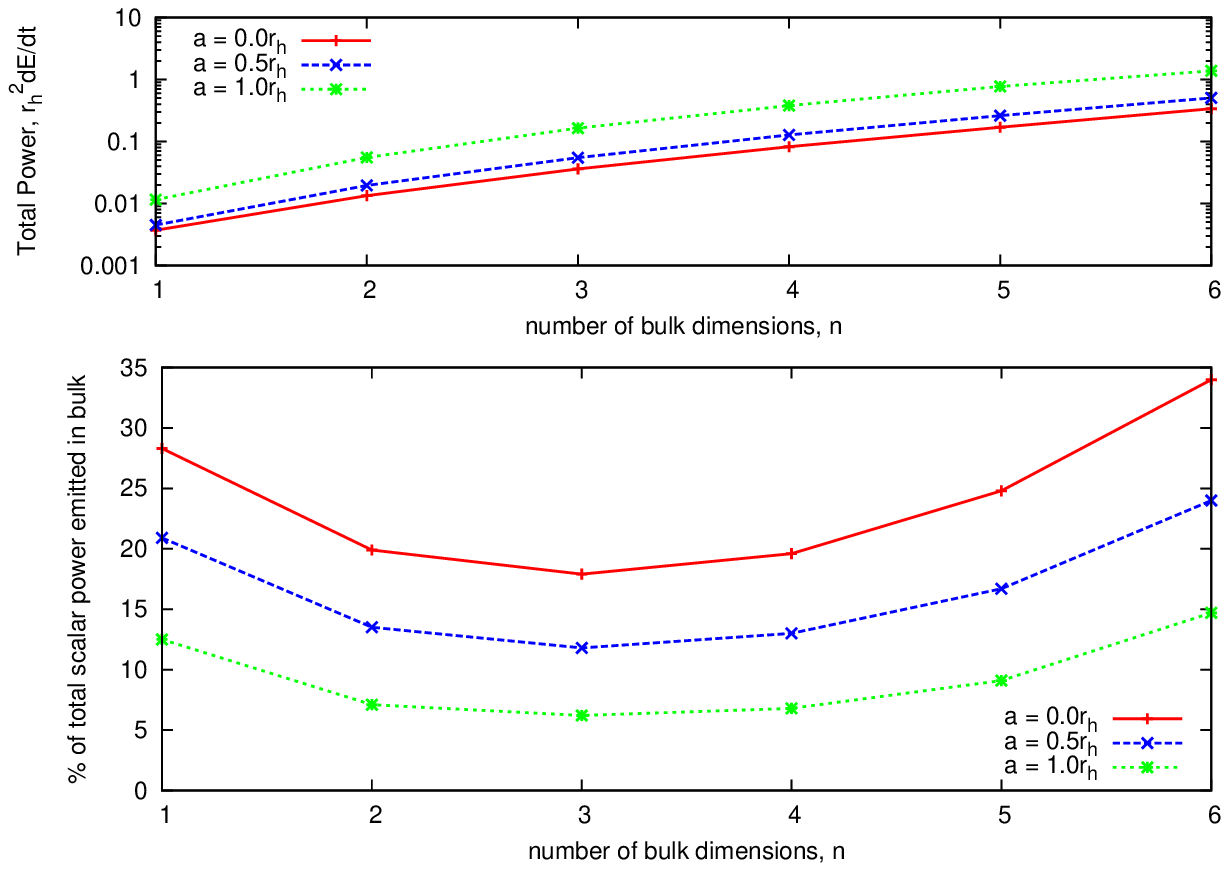}
%\end{center}
\caption{{\emph{Total scalar emission for a range of scenarios.}}
The upper plot shows the total power in brane and bulk scalars, for $\as = 0$, $0.5$ and $1$,
and for a range of numbers of extra dimensions, $n = 1,2 \ldots 6$.
The lower plot shows the $\%$ of this power which is emitted into the bulk. }
\label{fig-totals}
}

\TABLE{
%\begin{center}
\begin{tabular}{| c || c c | c c | c c |}
 \hline
 & \quad $\as = 0.0$ & & \quad $\as = 0.5$ & & \quad $\as = 1.0$ & \\
 \hline
 \quad $n = 1$ \quad & \quad $28.3\%$ of & $0.00371$   \quad & \quad $20.9\%$ of & $0.00449$ \quad & \quad $12.5\%$ of & $0.01154$ \quad  \\
 \quad $n = 2$ \quad & \quad $19.9\%$ of & $0.0134$  \quad & \quad $13.5\%$ of & $0.01953$ \quad &  \quad $7.1\%$ of & $0.05533$ \quad \\
  \quad$n = 3$ \quad & \quad $17.9\%$ of & $0.0362$ \quad & \quad $11.8\%$ of & $0.05497$ \quad &  \quad $6.2\%$ of & $0.1646$ \quad  \\
 \quad $n = 4$ \quad & \quad $19.6\%$ of & $0.0823$  \quad & \quad $13.0\%$ of & $0.1275$ \quad & \quad $6.8\%$ of & $0.3808$ \quad  \\
 \quad $n = 5$ \quad & \quad $24.8\%$ of & $0.170$  \quad & \quad $16.7\%$ of & $0.2609$ \quad & \quad $9.1\%$ of & $0.7709$ \quad  \\
 \quad $n = 6$ \quad & \quad $34.0\%$ of & $0.339$ \quad & \quad $24.0\%$ of & $0.5041$ \quad & \quad $14.7\%$ of & $1.3832$ \quad \\
 \hline
\end{tabular}
%\end{center}
\caption{{\emph{Proportion of total power emitted into the bulk.}}
This table gives the percentage of the total power (i.e. the
combined bulk and brane mass loss rate, in units of $1/r_h^2$)
that is emitted into the bulk, for a range of dimensionalities
$n$, and angular momenta $\as$. The total powers were calculated
by numerically integrating the power spectra up to cutoffs of
$\omega r_h = 10$, $12$ and $16$, for $\as = 0.0$, $0.5$ and
$1.0$, respectively.  }
\label{tbl-proportions}
}

Finally, let us briefly consider very fast rotation, $\as > 1$.
As we have seen, a large number of modes are excited by a
fast-rotating black hole. For instance,
figure \ref{fig-n6a1_5-modes} shows that modes up to $l \sim 30$
contribute to the emission spectrum of a 10D hole at $\as = 1.5$,
and that the power spectrum is significant at $\omega r_h \sim 15$
and beyond. Such spectra may give a misleading impression, since
the black hole cannot emit a pair of particles which are more
energetic than the black hole mass itself. For black holes created
at a collider, we would expect the black hole mass to be only a
few multiples of $M_\ast$. Many other physical effects may modify
the spectrum for $\omega r_h \gg 1$. A concerted effort is
currently underway to model the particle showers produced by black
hole decay using Monte Carlo event generators \cite{montecarlo}.

\FIGURE{
%\begin{center}
\includegraphics[width=14cm]{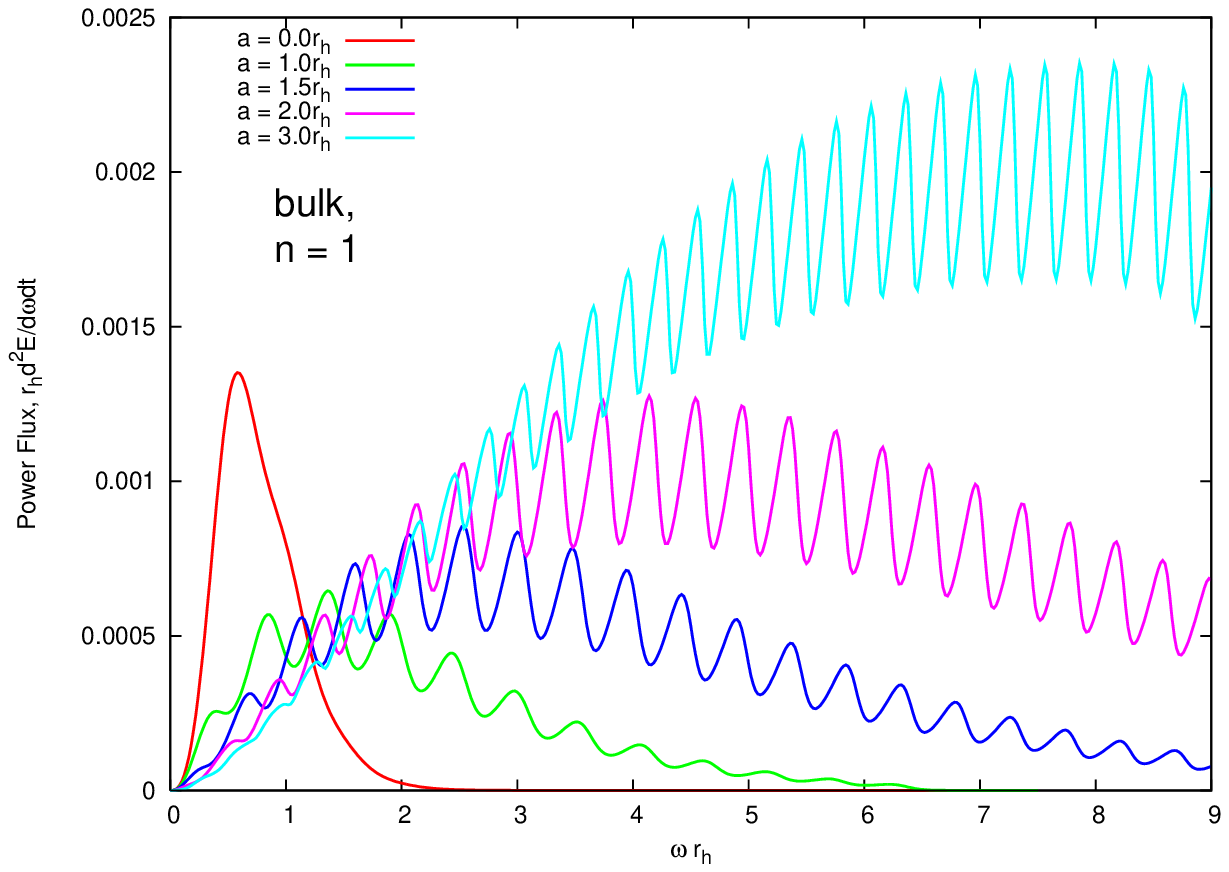}
%\end{center}
\caption{{\emph{Emission in the bulk from a fast-rotating 5D black hole.}}
The plot shows how the power emitted into the bulk varies with black hole rotation rate,
$\as = a/r_h = 0$, $1$, $1.5$, $2$ and $3$, with one extra dimension $n = 1$.}
\label{fig-high-a}
}

Regardless of these physical considerations, it is still possible
to compute an `idealised' semi-classical emission spectra using
equation (\ref{eq-power-emission}). The calculation is easiest at low
$n$, since many subdominant modes ($m < l - m_{\text{cutoff}}$)
can be neglected. Figure \ref{fig-high-a} shows bulk emission from
a 5D black hole at $\as = 0$, $1$, $1.5$, $2$ and $3$. Again, the
oscillatory structure is due to the dominance of the superradiant
($m = l$) modes. In every case we have studied we find the power
spectrum reaches an overall peak and tends to zero as $\omega r_h
\rightarrow \infty$.

\subsubsection{Angular momentum spectra}
A rotating black hole loses both mass and angular momentum through
the Hawking emission process.  The loss rate of angular momentum
is given by equation (\ref{eq-am}).

Figure \ref{fig-n2-a0to1-am} compares the loss rate of angular
momentum on the brane (left panel) and in the bulk (right panel),
for a range of rotations of a 6D ($n=2$) black hole. Note the
order-of-magnitude difference in their magnitudes. Hence, the
majority of angular momentum is emitted on the brane. This is
perhaps not surprising, since the plane of rotation lies on the
brane. Table \ref{table-ratios-n2-am} lists the total angular
momentum emission rates for $n = 2$, $\as = 0 \ldots 1.0$. The
proportion of angular momentum entering the bulk is found to
decrease monotonically with $\as$.

\FIGURE{
%\begin{center}
\includegraphics[width=14cm]{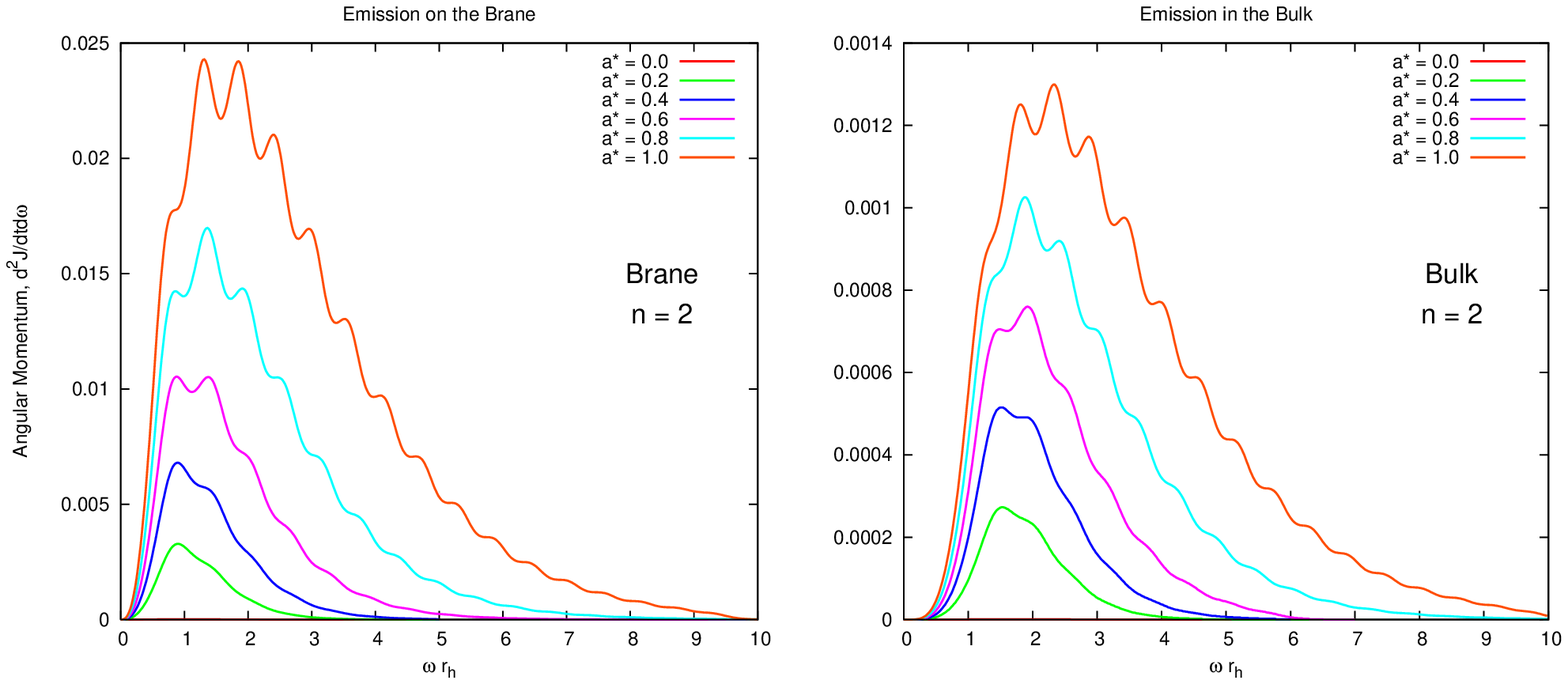}
%\end{center}
\caption{{\emph{Brane and bulk angular momentum from a 6D rotating black hole.}}
The left and right plots show the angular momentum emitted on the brane and in the bulk,
for a range of angular momenta $a_d = a / r_h$. Note the order-of-magnitude difference in the scales on the $y$-axis.}
\label{fig-n2-a0to1-am}
}

\TABLE{
%\begin{center}
\begin{tabular}{| l | c c c c c |}
\hline
 & $\as = 0.2  $ & $\as = 0.4 $ & $\as = 0.6 $ & $\as = 0.8 $ & $\as = 1.0 $ \\
\hline
Bulk A.M. & $ 4.04 \times 10^{-4} $ & $ 8.99 \times 10^{-4} $ & $ 1.60 \times 10^{-3} $ & $ 2.73 \times 10^{-3} $ & $ 4.54 \times 10^{-3} $ \\
Brane A.M. & $ 3.97 \times 10^{-3} $ & $ 9.98 \times 10^{-3} $ & $ 2.10 \times 10^{-2} $ & $ 4.17 \times 10^{-2} $ & $ 8.00 \times 10^{-2} $ \\
\% in Bulk & $ 9.2 \%$ & $ 8.3 \%$ & $ 7.1 \%$ & $ 6.1 \%$ & $ 5.4 \%$ \\
\hline
\end{tabular}
%\end{center}
\caption{{\emph{Proportion of (scalar) angular momentum emitted
into the bulk by a 6D rotating black hole.}}
These figures were calculated
by numerically integrating the angular momentum spectra, up to a
cutoff of $\omega r_h = 10$. The angular momentum loss rate is
given in units of $1/r_h$.}
\label{table-ratios-n2-am}
}

Figure \ref{fig-n6-a0to1-am} shows the angular momentum emitted by
a rotating black hole embedded in a ten-dimensional bulk.
Comparing with figure \ref{fig-n2-a0to1-am}, we see that the overall
emission rate has been enhanced by an order-of-magnitude by the
increase in bulk dimensionality. We estimated the total angular
momentum loss rate by integrating these spectra up to a cutoff of
$\omega r_h = 16$. We found that, at $\as = 0.5$, some $9.0\%$ of
the total angular momentum loss rate of $0.28 r_h^{-1}$ is emitted
in the bulk. At $\as = 1.0$, we estimate that $6.0\%$ of a total
of $1.64 r_h^{-1}$ is emitted in the bulk. Note that the latter
estimate is an underestimate due to the integration cutoff (see
figure  \ref{fig-n6-a0to1-am}).

%{\bf (In Figs.~\ref{fig-n2-a0to1-am} \&  \ref{fig-n6-a0to1-am} the curves for $\as =0$ are non-existent/non-seeable...?)}

\FIGURE{
%\begin{center}
\includegraphics[width=14cm]{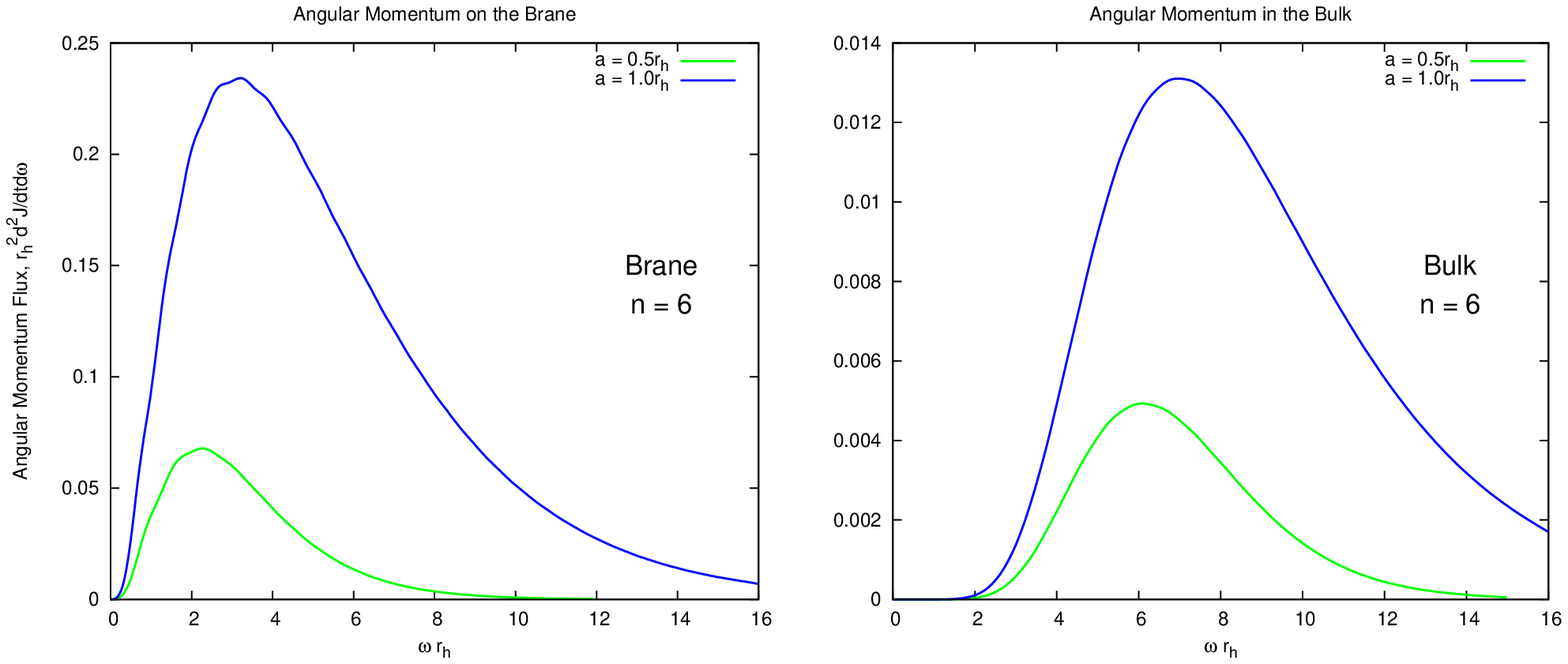}
%\end{center}
\caption{{\emph{Brane and bulk angular momentum from a 10D rotating black hole.}}
The left and right plots show the angular momentum emitted on the brane and in the bulk, respectively.
Note the order-of-magnitude difference in the scales on the $y$-axis.}
\label{fig-n6-a0to1-am}
}

\subsection{Consistency with other studies}

In this section we briefly compare our results with those
published in the literature.  We find our results are consistent
with the previous studies, where comparison is possible.
In all the papers mentioned below, the methods used to compute
the greybody factors are different from those employed here, and
therefore represent an independent check on our calculations.

First, let us consider the emission of scalars in the
Schwarzschild phase.  We find excellent agreement with the results
presented in \cite{HK1}, for both brane and bulk emission, although it should
be emphasized that the numerical algorithms used in \cite{HK1} are
different from those we have used in section \ref{sec-nummeth}.
%Compare, for example, the brane power spectra shown in figures
%\ref{schw-1}a, \ref{fig-n2-a0to1}a, \& \ref{fig-n6-a0to1}a with
%figure 2 in ~\cite{HK1}; and the bulk power spectra in figures
%\ref{schw-1}b \& \ref{fig-high-a} with figure 9 in~\cite{HK1}.
The bulk-to-brane proportions in table \ref{table-ratios-schw} agree
well with those in table 6 in \cite{HK1} (to $\sim 1$).  We also
find excellent agreement with the emission ratios presented in
table 1 of the second paper in \cite{graviton-schw}.

% Comparing with Cardoso, Cavaglia and Gualtieri
%            n = 0       1       2       3       4       5       6
% CCG ratio      1      8.94     36      99.8    222     429     749
% CDKW
% - brane pow 2.975e-4  0.00266  0.0107  0.0297  0.0662  0.1278  0.224
%  - ratio       1      8.94     36      99.8    222.5   429.6   752.9

Next, let us consider scalar emission on the brane in the rotating
phase; here, we are in agreement with \cite{DHKW}, which again
uses different numerical methodology to that in section \ref{sec-nummeth}.
%For example,
%compare the power spectra in figure \ref{fig-power-all-n}a with
%figures 6 \& 7 in~\cite{DHKW}; and the angular momentum spectrum in
%figures \ref{fig-n2-a0to1-am} \& \ref{fig-n6-a0to1-am} with figure 11
%in~\cite{DHKW}.
The total powers on the brane presented in table
\ref{tbl-proportions} are in reasonable agreement with the totals
in figure 12b in~\cite{DHKW}. Our values are likely to be more
accurate than those shown in figure 12b of \cite{DHKW}, because we
use a higher cutoff frequency ($\omega r_h \sim 16$ vs.~$\omega
r_h \sim 3$).

We have checked the credibility of our new results for scalar
emission in the bulk in a variety of ways.  For example, in section
\ref{sec-results} we confirmed that the numerical greybody factors
agree with the analytic approximation derived in \cite{CEKT} in the appropriate limit
($\omega r_h \lesssim l$). Comparison was also made with two
numerical studies for the special cases $n=1$ and $n=2$ (i.~e.~ 5D
and 6D black holes). A study of bulk emission from a 6D rotating
black hole lying on a tense brane was recently conducted in
\cite{kobayashi}. We confirmed with the authors that, when the
brane tension is set to zero, our power spectra are in excellent
agreement with their results (for the special cases $n=2$, $\as =
0$, $0.6$ and $1.2$). On the other hand, we cannot reproduce the
results of a recent study of a 5D hole \cite{jung-park-2005};
whilst we find good agreement for brane emission, we cannot verify
their results for bulk emission (figure 5, \cite{jung-park-2005}).
Since this appears to be the only point of disagreement between
our study and the existing literature, we remain confident that
our numerical results are substantively correct.

\section{Discussion and conclusions\label{sec-conclusions}}
The emission of Hawking radiation, in the form of elementary particles, by a
higher-dimensional black hole might be the most distinctive signature of the
creation of a black hole, and thus of the existence of extra dimensions in
nature. Whereas the emission on the brane may in principle be directly observable
and can be used as a source of valuable information about the extra
dimensions, the emission in the bulk -- the space transverse to our brane --
will only be interpreted as a missing energy signal. Obviously, the more
energy the black hole emits in the bulk, the less remains for emission
on the brane. On the other hand, the amount of missing energy, due to the
emission in the bulk and the emission of elusive particles such as neutrinos
on the brane, is now considered as another distinctive signature of a black
hole event. For the above reasons, the study of the emission in the bulk by
a higher-dimensional black hole is of equal importance as the one on the brane.

The bulk emission channel for a higher-dimensional black hole in its
spherically-symmetric Schwarzschild phase was the first one to be studied in
the literature. Two types of particles are usually considered as being
allowed to propagate in the bulk: gravitons and scalars. A comprehensive
study for the case of the latter type of particles \cite{HK1} demonstrated that
the emission in the bulk remains always subdominant to the one on the brane,
although it may become important for a large number of extra spacelike dimensions.
A number of studies in the case of gravitons \cite{graviton-schw} have finally
concluded that, although the emission of gravitons is enhanced by the existence
of extra dimensions more than any other species of particles, the total emission
of gravitational modes remains subdominant to the one corresponding to the
Standard Model degrees of freedom on the brane.

The study of the bulk emission during the spin-down phase of the life of the
black hole has up to now been restricted to scalar fields. This is due to the
increased complexity of the gravitational background around the rotating
higher-dimensional black hole, and to the cumbersome formalism necessary to
derive the equations of motion of higher-spin fields in the bulk. The existing
scalar analyses themselves, although very illuminating, are either approximate
\cite{CEKT} or refer to black holes with a specific dimensionality
\cite{jung-park-2005, kobayashi}. For this reason, here we have attempted to
present a comprehensive study of the bulk emission of scalar fields by a
rotating black hole with arbitrary angular momentum and living in a spacetime
with arbitrary number of extra dimensions.

The quantization of scalar fields propagating in a higher-dimensional, rotating
black-hole background closely follows the one for their 4-dimensional analogues,
therefore, in section 2 we have presented only the main steps of this analysis. It
is worth mentioning that well-known techniques \cite{F&T, Unruh'76, Christ'76, Christ'78,
 DeWitt, Schwinger, Synge} developed for 4-dimensional spacetime
are found to hold also in this case leading to similar expressions for the
energy and angular-momentum emission rates. Our numerical methods, used to
compute the angular eigenvalues and the transmission factors, were described
in detail in section 3. Finally, our exact numerical results for the bulk
scalar emission from a rotating black hole were presented in section 4.

As a consistency check, our numerical analysis first reproduced the previously
derived results for the emission of scalar fields by a non-rotating black hole
in the bulk \cite{HK1}. We then proceeded to study the rotating case,
and demonstrated that the approximate analysis presented in
\cite{CEKT} was in very good agreement with the exact numerical one in
the low-energy regime. Inevitably, deviations appear beyond this regime, necessitating a full numerical study.
The computation of the value of the transmission factor at all energies for
a rotating black hole demands the consideration of a very large number of
angular modes, that increases as the energy increases too. Our final results
for the transmission factor revealed a hierarchical splitting of the modes
according to the angular momentum numbers (first on $l$, then on $m$, and
finally on $j$) as well as the expected superradiance effect, that was most
effective for the ($m=l$, $j=0$) modes and for low values of $n$. Compared
to the non-rotating case, the transmission factor for the most-dominant
modes is found to be enhanced with the angular momentum of the black hole.

The corresponding power spectra for scalar fields were then computed. For
the purpose of comparing bulk and brane emission, we derived both types of
spectra for the same set of values of $a$ and $n$. We find that whereas
the brane scalar emission is enhanced with the angular momentum of the
black hole, in accordance with previous studies, the bulk emission is
initially suppressed and starts increasing only after a certain value
of $a$. This effect persists for all values of $n$ and is caused by
the fact that -- contrary to the emission on the brane -- the enhancement
of the transmission factor in the bulk is not in a position to compensate
for the decrease in the black-hole temperature with $a$. The complete picture emerges
from the entries of table 3, which lists the proportion of the total
power emitted by the black hole in the bulk. From these, we see that,
for all examined values of $a$ and $n$, the bulk scalar emission is only
a fraction of the brane emission. This fraction becomes important only for
large values of $n$, as in the non-rotating case, and even then it is
suppressed as $a$ increases. We may thus safely conclude that, as far
as the scalar channel is concerned, a rotating black hole emits a smaller fraction of its
mass into the bulk than a non-rotating one. Our results therefore seem to
add extra support to the argument presented in \cite{emparan} that black
holes radiate mainly on the brane.

Finally, we considered the angular-momentum loss rate of the black hole.
Perhaps not surprisingly since the black hole's sole angular momentum
component lies on the brane, we found again that the black hole loses its
angular momentum mainly by emission on the brane rather than in the bulk.
Thus, for the indicative case of a 6-dimensional black hole we found that
the proportion of the angular momentum emitted in the bulk can be
described with single figures, as can be seen in table 4.

The analysis of the brane-to-bulk energy balance is, however, not yet
complete. The graviton emission channel for the rotating phase needs to be investigated before
a final conclusion can be drawn. In the non-rotating case \cite{graviton-schw},
it was found that the behaviour of the transmission factors for gravitons
and scalar fields in the bulk shared a number of qualitative features, with
the significant enhancement of the graviton power rate caused mainly by
the larger number of degrees of freedom in the bulk. It remains to be seen whether, for a rotating black hole,
the graviton transmission factors will exhibit the same
characteristics as the scalar transmission factors found in this work. If so, it also remains to be seen whether
 the number of graviton states can tilt the balance towards
bulk emission. For this, we would need the perturbed equation of motion
for gravitons in a general rotating, higher-dimensional black-hole
background which is unfortunately still missing from the literature.

\acknowledgments
M.C. wishes to
acknowledge Science Foundation Ireland for financial support.
S.D.
acknowledges financial support from the Irish Research Council for
Science, Engineering and Technology (IRCSET).
P.K. and E.W. wish to thank  University College
Dublin for hospitality during the early stages of this work.
P.K. also acknowledges financial support from the UK STFC
PPA/A/S/2002/00350 research grant and participation in the
RTN networks UNIVERSENET-MRTN-CT-2006-035863-1 and MRTN-CT-2004-503369.
The work of E.W. was supported by UK STFC,
grant numbers PPA/G/S/2003/00082 and PPA/D000351/1.

\end{document}